\documentclass[reprint, aps, nofootinbib, prb]{revtex4-1}
\usepackage{amsmath} \usepackage{graphicx} \usepackage{dcolumn}
\usepackage{bm} 
\usepackage{amssymb}
\usepackage{pstricks}
\usepackage{subfigure}

\begin{document}

\newlength{\figurewidth}
\setlength{\figurewidth}{\columnwidth}

\newcommand{\prtl}{\partial}
\newcommand{\la}{\left\langle}
\newcommand{\ra}{\right\rangle}
\newcommand{\dla}{\la \! \! \! \la}
\newcommand{\dra}{\ra \! \! \! \ra}
\newcommand{\we}{\widetilde}
\newcommand{\smfp}{{\mbox{\scriptsize mfp}}}
\newcommand{\smp}{{\mbox{\scriptsize mp}}}
\newcommand{\sph}{{\mbox{\scriptsize ph}}}
\newcommand{\sinhom}{{\mbox{\scriptsize inhom}}}
\newcommand{\sneigh}{{\mbox{\scriptsize neigh}}}
\newcommand{\srlxn}{{\mbox{\scriptsize rlxn}}}
\newcommand{\svibr}{{\mbox{\scriptsize vibr}}}
\newcommand{\smicro}{{\mbox{\scriptsize micro}}}
\newcommand{\scoll}{{\mbox{\scriptsize coll}}}
\newcommand{\sattr}{{\mbox{\scriptsize attr}}}
\newcommand{\sth}{{\mbox{\scriptsize th}}}
\newcommand{\sauto}{{\mbox{\scriptsize auto}}}
\newcommand{\seq}{{\mbox{\scriptsize eq}}}
\newcommand{\teq}{{\mbox{\tiny eq}}}
\newcommand{\sinn}{{\mbox{\scriptsize in}}}
\newcommand{\suni}{{\mbox{\scriptsize uni}}}
\newcommand{\tin}{{\mbox{\tiny in}}}
\newcommand{\scr}{{\mbox{\scriptsize cr}}}
\newcommand{\tstring}{{\mbox{\tiny string}}}
\newcommand{\sperc}{{\mbox{\scriptsize perc}}}
\newcommand{\tperc}{{\mbox{\tiny perc}}}
\newcommand{\sstring}{{\mbox{\scriptsize string}}}
\newcommand{\stheor}{{\mbox{\scriptsize theor}}}
\newcommand{\sGS}{{\mbox{\scriptsize GS}}}
\newcommand{\sBP}{{\mbox{\scriptsize BP}}}
\newcommand{\sNMT}{{\mbox{\scriptsize NMT}}}
\newcommand{\sbulk}{{\mbox{\scriptsize bulk}}}
\newcommand{\tbulk}{{\mbox{\tiny bulk}}}
\newcommand{\sXtal}{{\mbox{\scriptsize Xtal}}}
\newcommand{\sliq}{{\text{\tiny liq}}}

\newcommand{\smin}{\text{min}}
\newcommand{\smax}{\text{max}}

\newcommand{\saX}{\text{\tiny aX}}
\newcommand{\slaX}{\text{l,{\tiny aX}}}

\newcommand{\svap}{{\mbox{\scriptsize vap}}}
\newcommand{\sjam}{J}
\newcommand{\Tm}{T_m}
\newcommand{\sTS}{{\mbox{\scriptsize TS}}}
\newcommand{\sDW}{{\mbox{\tiny DW}}}
\newcommand{\cN}{{\cal N}}
\newcommand{\cB}{{\cal B}}
\newcommand{\br}{\bm r}
\newcommand{\be}{\bm e}
\newcommand{\cH}{{\cal H}}
\newcommand{\cHlt}{\cH_{\mbox{\scriptsize lat}}}
\newcommand{\sthermo}{{\mbox{\scriptsize thermo}}}

\newcommand{\bu}{\bm u}
\newcommand{\bk}{\bm k}
\newcommand{\bX}{\bm X}
\newcommand{\bY}{\bm Y}
\newcommand{\bA}{\bm A}
\newcommand{\bb}{\bm b}

\newcommand{\lintf}{l_\text{intf}}

\newcommand{\DV}{\delta V_{12}}
\newcommand{\sout}{{\mbox{\scriptsize out}}}
\newcommand{\dv}{\Delta v_{1 \infty}}
\newcommand{\dvin}{\Delta v_{2 \infty}}

\newcommand*\xbar[1]{%
  \hbox{%
    \vbox{%
      \hrule height 0.5pt 
      \kern0.5ex
      \hbox{%
        \kern-0.1em
        \ensuremath{#1}%
        \kern-0.1em
      }%
    }%
  }%
}

\def\Xint#1{\mathchoice
   {\XXint\displaystyle\textstyle{#1}}%
   {\XXint\textstyle\scriptstyle{#1}}%
   {\XXint\scriptstyle\scriptscriptstyle{#1}}%
   {\XXint\scriptscriptstyle\scriptscriptstyle{#1}}%
   \!\int}
\def\XXint#1#2#3{{\setbox0=\hbox{$#1{#2#3}{\int}$}
     \vcenter{\hbox{$#2#3$}}\kern-.5\wd0}}
\def\ddashint{\Xint=}
\def\dashint{\Xint-}
\title{On the Mechanism of Activated Transport in Glassy Liquids}

\author{Vassiliy Lubchenko} \email{vas@uh.edu} \affiliation{Department
  of Chemistry, University of Houston, Houston, TX 77204-5003}
\affiliation{Department of Physics, University of Houston, Houston, TX
  77204-5005}

\author{Pyotr Rabochiy} \affiliation{Department of Chemistry,
  University of Houston, Houston, TX 77204-5003} 


\date{\today}

\begin{abstract}

  {\bf ABSTRACT:} We explore several potential issues that have been
  raised over the years regarding the ``entropic droplet'' scenario of
  activated transport in liquids, due to Wolynes and coworkers, with
  the aim of clarifying the status of various approximations of the
  random first order transition theory (RFOT) of the structural glass
  transition. In doing so, we estimate the mismatch penalty between
  alternative aperiodic structures, above the glass transition; the
  penalty is equal to the typical magnitude of free energy
  fluctuations in the liquid.  The resulting expressions for the
  activation barrier and the cooperativity length contain exclusively
  bulk, static properties; in their simplest form they contains only
  the bulk modulus and the configurational entropy per unit
  volume. The expressions are universal in that they do not depend
  explicitly on the molecular detail.  The predicted values for the
  barrier and cooperativity length and, in particular, the temperature
  dependence of the barrier are in satisfactory agreement with
  observation.  We thus confirm that the entropic droplet picture is
  indeed not only internally-consistent but is also fully
  constructive, consistent with the apparent success of its many
  quantitative predictions. A simple view of a glassy liquid as a
  locally metastable, degenerate pattern of frozen-in stress emerges
  in the present description. Finally, we derive testable
  relationships between the bulk modulus and several characteristics
  of glassy liquids and peculiarities in low-temperature glasses. \\

  {\em Keywords}: glass transition, supercooled liquids,
  $\alpha$-relaxation, RFOT theory, nucleation, configurational
  entropy

\end{abstract}

\maketitle


\section{Motivation}

The mechanism of activated transport in liquids near the glass
transition and its dramatic slowing down with compression and/or
cooling remains a subject of controversy.~\cite{0034-4885-77-4-042501,
  biroli:12A301} It is also of great interest in applications;
understanding the mechanism will allow us to reliably estimate the
relaxation barriers in liquids and thus make predictions on the
glass-forming ability of specific substances, among many other things.

The present article attempts to clarify several aspects of the
``entropic droplet'' mechanism of the activated transport, due to
Wolynes and coworkers,~\cite{KTW, MCT1, XW, LW_aging} and obtains
simple, qualitative expressions for the activation barrier and the
cooperativity length for the $\alpha$-relaxation that connect to
material properties very explicitly. The argument also delineates the
role of local fluctuations in the energetics of the activated
transport.

According to the random first order transition (RFOT) theory, liquids
undergo a crossover from mainly collisional to activated transport at
sufficiently high densities and/or low temperatures.\cite{LW_soft} A
thermodynamic signature of the crossover is that the liquid density
profile is no longer uniform but, instead, becomes a collection of
sharp peaks.~\cite{dens_F1, dens_F2, BausColot, Lowen, RL_LJ,
  PhysRevLett.82.747} The peaks correspond to particles vibrating
around certain positions in space, analogously to regular, periodic
crystals. In contrast with the periodic crystals, the aperiodic
structures are however {\em transient}, even if long-lived. Mutual
reconfiguration between distinct, long-lived aperiodic structures
occurs via local activated events.  As a result of these local
reconfigurations, the liquid flows and the translational symmetry is
eventually restored.  In kinetic terms, the crossover to activated
transport corresponds to a well-developed timescale separation between
atomic vibrations and translations.~\cite{LW_soft, LW_Wiley}
Appropriately, in the meanfield limit the crossover corresponds with
the kinetic catastrophe of the mode-coupling theory,~\cite{MCT}
whereby all translations freeze completely.

Both the long-lived structures and the mutual reconfigurations have
been observed, respectively, by neutron scattering~\cite{MezeiRussina}
and a variety of non-linear spectroscopic
methods.\cite{GruebeleSurface, Spiess, RusselIsraeloff,
  CiceroneEdiger} The crossover to the activated transport may occur
both above or below the fusion temperature, the two extremes
corresponding to very strong and very fragile liquids
respectively.~\cite{LW_soft} In the latter case, a liquid below the
crossover is formally supercooled. For generality, we shall call
liquids in the activated transport regime {\em glassy}, so as to
include in the analysis strong liquids that are already in the
activated regime but are not technically supercooled.

In the specific scenario by Wolynes and coworkers,\cite{KTW, MCT1,
  XW, LW_aging} the activated reconfigurations proceed by a mechanism
akin to {\em nucleation}, the corresponding free energy profile given
by:
\begin{equation} \label{FR} F(R) = \frac{4\pi}{3} R^3 \Delta \tilde{f}
  + 4\pi R^2 \sigma(R).
\end{equation}
The bulk driving force $\Delta \tilde{f}$, in equilibrium, is entirely
due to multiplicity of alternative aperiodic states: $\Delta \tilde{f}
= - T \tilde{s}_c$, where $\tilde{s}_c$ is the configurational entropy
per unit volume.  Lubchenko and Wolynes~\cite{LW_aging} have clarified
this notion using a library construction of aperiodic liquid states,
which explicitly shows that although an individual aperiodic state is
replaced by an individual state, the reconfiguration itself is still
driven by the configurational entropy, which reflects the full
ensemble of thermally relevant, distinct aperiodic states. The
configurational entropy is the log-number of such thermally relevant
states and is a convenient measure of the degeneracy of the liquid
free energy landscape. Counting such aperiodic states becomes
unambiguous below the crossover, when particle translations and
vibrations are timescale-separated and so mutual reconfigurations
between the states are technically {\em rare events}. The rate of such
rare events can be evaluated quantitatively using the transition state
theory.~\cite{Kramers, FW, Hanggi_RMP}

The quantity $4\pi R^2 \sigma(R)$ has the formal structure of a
surface tension term and accounts for the mismatch penalty between
alternative aperiodic structures. Kirkpatrick, Thirumalai, and
Wolynes~\cite{KTW} (KTW) pointed out that the free energies of the
aperiodic structures are {\em distributed} implying that a smooth
interface between two such structures would generally distort somewhat
to minimize the local free energy. This is because the penalty for
small increases in the interface area increases slower with the extent
of the deformation than the concomitant stabilization of the {\em
  bulk} free energy. KTW noticed that this situation is analogous to
the random-field Ising model (RFIM), in which the Zeeman splittings on
separate sites are uncorrelated Gaussian random variables, where the
distribution width is $h$. The total field in a region of size $N$
thus scales as $h \sqrt{N}$. Hereby, a smooth interface between
spin-up and down domains would distort to stabilize the randomly
distributed Zeeman energy. Villain~\cite{Villain} devised a
coarse-graining procedure by which one can estimate the effective
surface tension coefficient $\sigma(R)$, in the RFIM, given the
curvature $1/R$ of the interface. Assuming that the hyperscaling
relation $\alpha = 2 - \nu D$ (Ref.~\onlinecite{Goldenfeld}) for the
heat capacity and the correlation length holds in the large droplet
limit, KTW~\cite{KTW} have fixed the boundary condition for the
$\sigma(R)$ renormalization at infinity, $\sigma(\infty)=0$, to arrive
at a particularly simple, scale-free relation between the effective
surface tension coefficient and the interface curvature:
\begin{equation} \label{sigmaR}
\sigma(R) = \sigma_0 (a/R)^{1/2}, 
\end{equation}
where the surface tension coefficient $\sigma_0$ at the molecular
scale $a$ is proportional to the field $h$:
\begin{equation} \label{sigmah} \sigma_0 \propto h.
\end{equation}
Xia and Wolynes~\cite{XW} (XW) used notions of the classical
density-functional theory (DFT) to estimate $\sigma_0$ without using
adjustable parameters. This estimate leads to dozens of quantitative
predictions for distinct glassy phenomena, both classical and quantum,
see Refs.~\onlinecite{LW_ARPC, LW_RMP, L_JPCL} for reviews and
Refs.~\onlinecite{Wisitsorasak02102012, RL_sigma0, RWLbarrier,
  PhysRevE.88.022308} for subsequent work. Recently, the surface
tension coefficient $\sigma_0$ has been estimated using standard DFT
methods,~\cite{RL_sigma0} the results consistent with the simpler XW
argument. Thus the original fluctuation field $h$, which was
introduced by KTW seemingly phenomenologically, would appear to be
determined retroactively.

The nuclei of incipient aperiodic states, which evolve according to
Eq.~(\ref{FR}), have been called {\em entropic droplets}~\cite{MCT1,
  KTW} to reflect the entropic nature of the driving force for the
nucleation.  The critical size $R^\ddagger$ for an entropic droplet,
at which $\prtl F/\prtl R = 0$, is given by
\begin{equation} \label{Rsc} R^\ddagger = \left(3 \sigma_0 a^{1/2}/2 T
    \tilde{s}_c \right)^{2/3},
\end{equation}
while the barrier itself is 
\begin{equation} \label{FRsc} F^\ddagger \equiv F(R^\ddagger) = 3 \pi
  \sigma_0^2 a/T \tilde{s}_c = 2 \pi \sigma_0 a^{1/2}
  (R^\ddagger)^{3/2} .
\end{equation}
This allows one to write down a simple relationship between the
relaxation time and the critical radius:
\begin{equation} \label{tauR} \tau = \tau_0 \exp\left(
     \frac{2 \pi \sigma_0 a^{1/2}}{k_B T} \, {R^\ddagger}^{3/2} \right).
\end{equation}
If the aforementioned hyperscaling relation were not assumed to be
obeyed, the exponent $3/2$ would be replaced by $2$, however the
relation between $\log \tau$ and $R^\ddagger$ is still robustly
scale-free within the nucleation scenario.

Despite the apparent quantitative success of the entropic droplet
picture, the latter is far from being universally accepted, see for
instance Refs.~\onlinecite{0034-4885-77-4-042501,
  biroli:12A301}. Several aspects of the scenario appear to be
particularly confusing or controversial, which we highlight in what
follows.

The free energies of the liquid before and after reconfiguration are
on average equal, prompting a question as to what exactly drives the
reconfiguration. The entropic driving force in Eq.~(\ref{FR}) seems
{\em non-standard} from the viewpoint of traditional nucleation
theory. For instance, because the bulk driving force is entirely
entropic while there is an energetic component to the mismatch
penalty---in systems other than strictly rigid particles---it may
appear that the energy of the system {\em increases} after each
nucleation event.  Clearly this cannot be true in equilibrium in view
of energy conservation.

Furthermore, nucleation profile (\ref{FR}) suggests, at a first
glance, that the nucleation event will proceed indefinitely toward $R
\to \infty$, implying the reconfigurations are not local. Xia and
Wolynes~\cite{XW} and others~\cite{LW} have argued the reconfiguration
will be completed when the droplet size reaches a value $R^*$ such
that $F(R^*) = 0$, which is the value of $F(R)$ prior to the
reconfiguration event. The length $R^*$ thus gives the size of a
cooperative region; consequently one may think of a glassy liquid as a
{\em mosaic} of regions that are relatively stabilized and are
physically separated by relatively strained regions that are similar
to interfaces between coexisting phases. (There are also important
distinctions, to be discussed below.) Since $R^* = 2^{2/3}
R^\ddagger$, one obtains, by Eq.~(\ref{tauR}), an asymptotic scaling
$R^* \propto (\log \tau)^{2/3}$ which seems to agree with
state-of-the-art simulations on a variety of liquid
models~\cite{PhysRevLett.112.097801} at the lowest accessed
temperatures although generally disagrees with higher temperature
data. The latter disagreement is expected since at increasing
temperature, activated dynamics are progressively affected by
mode-coupling and barrier-softening effects.~\cite{LW_soft, LW_Wiley}
The identification of $R^*$ as the equilibrium droplet size implies,
however, that the free energy $F(R)$ does not reach a minimum at
equilibrium and thus is distinct from the free energy of the system,
which begs for further clarification.

In addition, because the fluctuation field $h$ is not determined at
the onset of the calculation of the activation barrier---even though
it can be eventually estimated---the calculation of the mismatch
penalty in Eq.~(\ref{FR}) might be regarded as not entirely
constructive, even if internally-consistent. Related to this is a
simple notion that if there is no field $h$ to begin with, $h=0$, one
may superficially deduce that $\sigma = 0$, by Eqs.~(\ref{sigmaR}) and
(\ref{sigmah}), which is clearly not true since there must be a
mismatch penalty between distinct free energy minima. Perhaps for
these reasons, many have regarded the nucleation scenario largely as
an analogy to the random field Ising model,~\cite{PhysRevE.86.052501}
not a self-contained microscopic picture. Even more extremely, the
nucleation picture of the activated transport is often
regarded~\cite{0034-4885-77-4-042501, BouchaudBiroli} as a variation
on the heuristic arguments of Adam and Gibbs.~\cite{AdamGibbs}
  
Bouchaud and Biroli\cite{BouchaudBiroli} (BB) have put forth a
somewhat different view of the activated transport, though still
within the landscape paradigm. In this view, the ensemble of all
states of a compact region of size $\xi$ consists of a contribution
from the current state and contributions of the full, exponentially
large set of alternative structures.  As in the library construction,
the surrounding of the chosen region is constrained to be static up to
vibrational displacement. What sets apart the current state from all
the alternative states is that it fits the environment better.  The
mismatch energy must scale with the region size $N$ sublinearly while
the log-number of alternative states scales linearly. Thus the
stability of sufficiently {\em small} regions---which are smaller than
a certain length $\xi$ analogous to $R^*$---can be understood
thermodynamically in a straightforward manner: The energetic advantage
of being in the current state, due to the matching boundary, outweighs
the multiplicity of poorer matching, higher energy states. (This is
not unlike the stability of a crystal relative to the liquid below
freezing.)

The just listed aspects of the Bouchaud-Biroli picture are essentially
equivalent to the library construction, and, in particular, with
regard to the entropic nature of the driving force for the activated
transport.  In contrast with the KTW~\cite{KTW} and library
construction,~\cite{LW_aging} however, the BB scenario is agnostic as
to the concrete mechanism of mutual reconfiguration between
alternative aperiodic states, other that the reconfigurations must be
rare, activated events.  In the absence of such a concrete mechanism,
BB end up assuming a generic scaling relation between the
cooperativity length $\xi$ and relaxation time that is similar to
Eq.~(\ref{tauR}); the exponent is generally different from $3/2$, but
necessarily {\em positive}. In this view, a region {\em larger} than
the size $R^*$ can still reconfigure via a {\em single} activated
event but would do so typically {\em more slowly} than the region of
size $R^*$. Combining this notion with the lower bound on the
cooperativity size obtained above one concludes that the cooperativity
size is in fact $R^*$ and one does not face the subtlety stemming from
the downhill decrease of the free energy profile $F(R)$ from
Eq.~(\ref{FR}).

Non-withstanding its elegance, the Bouchaud-Biroli picture does rely
on scale-free relations similar to Eqs.~(\ref{Rsc}) and (\ref{tauR}),
which is not necessarily innocuous. Such scale-free relations would
naturally arise if the lengthscale $R^\ddagger$ (and $R^* \propto
R^*$) could in principle diverge, which is only possible if the
configurational entropy $s_c$ could vanish, by Eq.~(\ref{Rsc}). (To
avoid confusion, we stress that such vanishing could not be observed
in practice, by Eq.~(\ref{FRsc}).) Remarkably, the configurational
entropy extrapolated beyond the glass transition temperature would
vanish at a {\em finite} temperature often denoted as $T_K$.  The
vanishing of the configurational entropy---which is often referred to
as the Kauzmann crisis~\cite{Kauzmann}---has been one of the most
disputed topics in the glass transition field. There are mean-field
models, such as the celebrated Potts glass,~\cite{MCT1} which exhibits
both a kinetic catastrophe analogous to the crossover from the
collisional to activated transport, and an entropy crisis whereby the
degeneracy of the landscape vanishes at a finite temperature. Yet the
applicability of mean-field Potts models to liquids has been
questioned.~\cite{0034-4885-77-4-042501} Note however that two recent
works have shown finite-dimensional Potts-like models could in fact
exhibit the RFOT transition.~\cite{2014arXiv1407.7393C,
  2014arXiv1408.1495T} Additionally, spin models have been analyzed
that exhibit the kinetic and thermodynamic catastrophes in mean-field,
but in finite dimensions, the Kauzmann crisis disappears and no
diverging length of the type in Eq.~(\ref{Rsc}) is
found.~\cite{MooreNOdiverge, PhysRevE.86.052501, PhysRevB.85.100405}
Last, but not least, Stevenson and Wolynes~\cite{SWultimateFate} have
argued that a supercooled liquid will crystallize or partially order
before the configurational entropy could vanish.

We stress that in the nucleation scenario, formula (\ref{Rsc})---which
gratuitously happens to be also a scaling relation---is {\em derived}
on a microscopic basis; it is not postulated and does not rely in any
way on the vanishing of the configurational entropy. For this reason,
comparisons with spin models that may or may not experience the
Kauzmann crisis are not consequential so long as that the nucleation
mechanism is in fact correct.  Now, because the KTW and subsequent
derivations are quite specific as to the details of the activated
event and the value of the surface tension, the nucleation scenario
allows one to make not only qualitative, but also, apparently, {\em
  quantitative} predictions for many, seemingly disparate phenomena as
already mentioned. To establish whether these successful predictions
are not mere coincidences, it is imperative to ascertain whether or
not the entropic-droplet picture is in fact fully constructive.

The present work clarifies, we believe, the potential issues listed
above, which have been raised with regard to the entropic droplet
scenario over the years. We thus confirm the constructive nature of
the corresponding microscopic description.  The key notion that
resolves all of those seeming paradoxes is that, on the one hand, the
free energies of the individual free energy minima are distributed. On
the other hand, the minima cannot interconvert by means other than the
activated reconfigurations themselves. This is in contrast with
nucleation of a minority-phase~\cite{Bray} during regular first order
transitions, in which the microstates within individual phases
interconvert on time scales much {\em shorter} than the duration of
the nucleation event. That the free-energy fluctuations in glassy
liquids are frozen-in on the time scale of the nucleation event makes
all the difference.  As part of the argument, we show directly that
the original KTW prescription that the local field $h$ be determined
by the magnitude of free energy fluctuations is straightforwardly
implemented using a standard calculation. The resulting expression for
the activation barrier for liquid transport is very simple and
confirms the physical view of a glassy liquid as a structurally
degenerate pattern of frozen-in free energy
fluctuations.~\cite{BL_6Spin} We believe this expression represents a
good zeroth-order estimate of the barrier that possesses a great deal
of universality as it boils down to only two parameters reflecting the
molecular interactions, viz., the bulk modulus and configurational
entropy.  Despite its qualitative nature, the argument produces
barrier values that are in reasonable agreement with experiment.

The paper is organized as follows: Section~\ref{mosaic} discusses the
entropic driving force while Section~\ref{mismatch} revisits the
argument for the mismatch penalty. We test the expressions obtained
for the barrier against the experiment in Section~\ref{tests}. In
Section~\ref{conclusions}, we summarize, derive several scaling
relations for glassy liquids, and discuss implications of the present
results for certain anomalies observed in cryogenic glasses.

\section{Glassy liquid as a mosaic made of entropic droplets}
\label{mosaic}

We are used to systems whose free energy surface is essentially
independent of the system size: For instance, the free energy surface
of a macroscopic Ising ferromagnet below its Curie point has two
distinct free energy minima that can be distinguished by their average
magnetization, which is up or down respectively. If the system is made
thrice bigger, the free energy is simply multiplied by a factor of
three; that is, while the number of states within each minimum
increases exponentially, the number of minima themselves remains the
same, i.e., two.  In contrast, the number of free energy minima in a
glassy liquid scales exponentially with the system size $N$: $e^{s_c
  N}$, where $s_c$ is the configurational entropy per particle $s_c
\equiv \tilde{s}_c a^3$. Under these circumstances, the system will
break up into separate, contiguous regions that are relatively
stabilized; the contiguous regions are separated by relatively
strained interfaces characterized by a higher free energy density. To
see this, suppose the opposite were true and the free energy density
were uniform throughout. Owing to the multiplicity of distinct free
energy minima, the nucleation rate for another relatively stabilized
configuration is finite, as we will see shortly. As a result, the
original configuration will be {\em locally} replaced by another
configuration while the boundary of the replaced region will be
relatively strained because of a mismatch between the new structure
and its environment.  Thus in equilibrium, there is a steady-state
concentration of the strained regions; local reconfiguration takes
place at a steady rate between distinct aperiodic structures.  The
concentration of the strained regions and the escape rate from the
current liquid configuration can be determined self-consistently, as
discussed by Kirkpatrick, Thirumalai, and Wolynes,~\cite{KTW} Xia and
Wolynes,~\cite{XW} and Lubchenko and Wolynes.~\cite{LW_aging}

Here we provide a ``microcanonical'' version of that argument, which,
we hope, will make certain aspects of the nucleation scenario less
confusing. We start with the general expression for the partition
function of a thermodynamic system in contact with a thermal bath at
temperature $T \equiv 1/k_B \beta$ and pressure $p$:~\cite{McQuarrie,
  LLstat}
\begin{equation} \label{Z1} Z = C \int dE dV e^{-\beta[E + pV - TS(E,
    V)]}
\end{equation}
where $S(E, V)$ is the full entropy of the system as a function of
energy $E$ and volume $V$. $C$ is the normalization factor, see below.
In the usual manner, the integrand is very small for the smallest
values of the energy and volume because of the small number of the
corresponding configurations $e^{S(E, V)/k_B}$; it is also very small
for sufficiently large values of these arguments because of the
Boltzmann factor $e^{-\beta(E + pV)}$.  For a large enough system size
$N$, the integral is dominated by the vicinity of the integrand's
maximum at $E = \overline{E}$, $V = \overline{V}$---which thus
correspond to the most likely values of the respective
variables---while the probability distribution of the thermodynamic
variables is approximately Gaussian:
\begin{eqnarray} \label{Z2} Z & = & \int \frac{dE dV}{2 \pi |\prtl^2
    S(E, V)|^{-1}} e^{-\beta[\overline{E} + p\overline{V} -
    TS(\overline{E}, \overline{V})]} \\ & \times & \exp\left[
    \frac{1}{k_B} \left( \frac{1}{2} \frac{\prtl^2 S}{\prtl E^2}
      \Delta E^2 + \frac{\prtl^2 S}{\prtl E \prtl V} \Delta E \Delta V
      + \frac{1}{2} \frac{\prtl^2 S}{\prtl V^2} \Delta V^2 \right)
  \right], \nonumber
\end{eqnarray}
assuming a well-behaved $S(E, V)$ near $E = \overline{E}$, $V =
\overline{V}$.  Here, $\Delta E \equiv E - \overline{E}$, $\Delta V
\equiv V - \overline{V}$, and $|\prtl^2 S(E, V)|$ is the determinant
of the matrix of the second-order derivatives of the entropy with
respect to the energy and volume from the argument of the second
exponential on the r.h.s.  The derivatives are computed at $E =
\overline{E}$, $V = \overline{V}$. The first exponential in
Eq.~(\ref{Z2}) is independent of the integration variables and is
equal to $e^{-\beta \overline{G}}$, where $\overline{G} \equiv
\overline{E} + p\overline{V} - TS(\overline{E}, \overline{V})) \equiv
\overline{H} - TS(\overline{E}, \overline{V}))$ is the equilibrium
Gibbs free energy and $\overline{H}$ the enthalpy, of course.  By a
linear change of variables in the integral, one can determine the
magnitude of fluctuation of any thermodynamic quantity of interest; an
efficient way to do so is described in Chapter 112 of
Ref.~\onlinecite{LLstat}.  Here, we are specifically interested in the
fluctuation of the Gibbs free energy:
\begin{equation} \label{Z3} Z = \int \frac{dG}{\sqrt{ 2 \pi \delta G^2
    }} e^{-\beta \overline{G}} \, e^{-(G - \overline{G})^2/2 \delta
    G^2}.
\end{equation}
where $\delta G = \la (G - \overline{G})^2 \ra^{1/2}$. As shown in the
Appendix,
\begin{equation} \label{dG} \delta G = N^{1/2} \left[ \frac{k_B T
      K}{\bar{\rho}} + (K \alpha_t - \tilde{s})^2 \frac{k_B
      T^2}{\bar{\rho} \tilde{c}_v} \right]^{1/2},
\end{equation}
where $K \equiv -V (\prtl p/\prtl V)_T$ is the bulk modulus,
$\bar{\rho} \equiv 1/a^3$ average particle density, and $\alpha \equiv
(1/V) (\prtl V/\prtl T)_p$ thermal expansion coefficient. The
quantities $\tilde{c}_v$ and $\tilde{s}$ are, respectively, the heat
capacity at constant volume and entropy, both per unit volume.

Let us now consider a liquid below the crossover to activated
transport but above the glass transition; the liquid is thus
equilibrated. Below the crossover, the reconfigurations are rare
events compared with vibrational relaxation, which amounts to a
well-developed time scale separation between translations and
vibrations. This time-scale separation takes place in ordinary liquids
at viscosities of order $10$~Ps.~\cite{LW_soft, LW_Wiley} Because of
it, the entropy of the liquid can be written as a sum of distinct
contributions:
\begin{eqnarray} \overline{G} &=& \overline{H}_i - T
  \overline{S}_{\svibr, \, i} - T S_c \\ \label{gequil} &\equiv&
  \overline{G}_i - T S_c(\overline{G}_i), 
\end{eqnarray}
where $H_i$ is the enthalpy of an individual aperiodic state per
particle, while the total entropy is presented here as the sum of the
vibrational and configurational contributions, the configurational
contribution taking care of particle translations. The subscript
``$i$'' refers to ``individual'' metastable aperiodic states. The
quantity $\overline{G}_i \equiv \overline{H}_i - T
\overline{S}_{\svibr, \, i}$ would be the Gibbs free energy of the
sample if the particles were not allowed to reconfigure but were
allowed to vibrate only.  It is interesting that the free energy of
liquid was written in the form similar to Eq.~(\ref{gequil}) already
in 1937 by Bernal,~\cite{Bernal1937} who apparently assumed that
molecular vibrations and translations were distinct motions at {\em
  any} temperature, even though this notion is well-justified only
below the crossover.

Of direct interest is the distribution not of the full free energy $G$
but that of the free energy $G_i$ of individual metastable states:
\begin{equation} \label{Z3a} Z = \int \frac{dG_i}{\sqrt{ 2 \pi \delta
      G_i^2 }} e^{S_c(\overline{G}_i)/k_B-\beta \overline{G}_i} \,
  e^{-(G_i - \overline{G}_i)^2/2 \delta G_i^2}.
\end{equation}
The corresponding width of the distribution, $\delta G _i \equiv \la
(G_i - \overline{G}_i)^2 \ra^{1/2}$, can be evaluated similarly to
$\delta G$ from Eq.~(\ref{dG}), see the Appendix:
\begin{eqnarray} \label{dGi} \delta G_i = &N^{1/2} & \left\{ \left[ K
      - T \left(\frac{\prtl S_c}{\prtl V} \right)_T \right]^2
    \frac{k_B T}{K \bar{\rho}} \right. \nonumber \\ &+& \left. [K
    \alpha_t + ( \Delta \tilde{c}_v - \tilde{s}_\svibr) ]^2 \frac{k_B
      T^2}{\bar{\rho} \tilde{c}_v} \right\}^{1/2},
\end{eqnarray}
where $\Delta \tilde{c}_v \equiv T(\prtl \tilde{s}_c/\prtl T)_V$ is
the configurational heat capacity at constant volume and
$\tilde{s}_\svibr$ the vibrational entropy, both per unit volume.

It is convenient, for the present purposes, to shift the energy
reference so that $\overline{G}_i = 0$:
\begin{equation} \label{Z4} Z = \int \frac{dG_i}{\sqrt{ 2 \pi \delta
      G_i^2 }} e^{S_c/k_B} e^{-G_i^2/2 \delta G_i^2}.
\end{equation}
This way, the partition function gives exactly the number
$e^{S_c/k_B}$ of the (thermally available) states that is not weighted
by the Boltzmann factor $e^{-\beta \overline{G}_i}$.

Consider now a local region that is currently {\em not} undergoing a
structural reconfiguration.  Because the region is certainly known not
to be reconfiguring, its free energy---up to finite-size
corrections---is equal to $G_i$, which is typically higher than the
equilibrium free energy $\overline{G}$ from Eq.~(\ref{gequil}). The
free energy difference $\overline{G} - \overline{G}_i = - T S_c < 0$
is the driving force for the eventual escape from the current
structure, and, hence, relaxation toward equilibrium.  Next we
estimate the actual rate of escape and the typical region size that
will have reconfigured as a result of the escape event.

We specifically consider escape events that are local. Therefore, the
environment of a chosen compact region is static, up to vibration.
Consider the partition function for a compact region of size $N$
surrounded by such a static, aperiodic lattice. The vast majority of
the configurations do not fit the region's boundary as well as the
original configuration, and so there is a free energy penalty
$\Gamma_i > 0$ due to the mismatch between the static boundary and any
configuration of the region other than the original configuration.  We
anticipate that since local replacement of a structure amounts to a
legitimate fluctuation, $\Gamma$ and $\delta G_i$ should be
intrinsically related, which will indeed turn out to be the case.

In the presence of the mismatch penalty, the density of states can be
obtained by replacing $\overline{G}_i \to \overline{G}_i + \Gamma$
under the integral in Eq.~(\ref{Z3a}), where $\Gamma \equiv
\overline{\Gamma}_i$ is the typical value of the mismatch. The latter
generally scales with the region size:
\begin{equation} \label{Gamma1} \Gamma = \gamma N^x,
\end{equation}
but in a sub-thermodynamic fashion: $x < 1$, where the coefficient
$\gamma(N \to \infty) = \text{const}$.  Thus we obtain for the total
number of thermally available states for a region embedded in a static
lattice:
\begin{equation} \label{Z5} Z = \int \frac{dG_i}{\sqrt{ 2 \pi \delta
      G_i^2 }} e^{S_c/k_B - \beta \Gamma} e^{-(G_i-\Gamma)^2/2 \delta
    G_i^2},
\end{equation}
where we set the expectation value of the free energy in the {\em
  absence} of the penalty at zero, as before. (The expectation value
of $G_i$ corresponding to Eq.~(\ref{Z5}) is {\em not} zero at $N >
0$.) Note the argument of the first exponential on the r.h.s. is
independent of $G_i$ but does depend on the region size $N$, and so
does the total number of thermally available states $Z$:
\begin{equation} \label{Z6} Z(N) = e^{s_c N/k_B - \beta \gamma N^x},
\end{equation}
where $s_c \equiv S_c/N$ is the configurational entropy per particle.

Because of the sub-linear $N$-dependence of the mismatch penalty, the
number of thermally available states $Z(N)$ depends non-monotonically
on the region size.  For small values of $N$, this number {\em
  decreases} with the region size, which is expected since the region
is stable with respect to weak deformation such as movement of a few
particles. At the value $N^\ddagger$ such that $(\prtl Z/\prtl
N)_{N^\ddagger} = 0$, the number of available stats reaches its
smallest value and increases with $N$ for all $N > N^\ddagger$. This
critical size $N^\ddagger$:
\begin{equation} \label{Ncr1} N^\ddagger = \left(\frac{x \gamma}{T
      s_c}\right)^{\frac{1}{1-x}},
\end{equation}
corresponds to the least likely size of a rearranging region, and thus
corresponds to a bottleneck configuration for the escape event:
Indeed, any state at $N < N^\ddagger$ is less likely than the initial
state and so cannot be a final state upon a reconfiguration; such
final state must thus be at $N > N^\ddagger$. On the other hand, to
move any number of particles $N$ in excess of $N^\ddagger$, one must
have moved $N^\ddagger$ particles as an intermediate step.

The size $N^* > N^\ddagger$ such that
\begin{equation} \label{Z*} Z(N^*) = 1
\end{equation}
is special in that the region of this size has a thermally available
configuration, other than the original one, even though the boundary
is fixed. By construction, this configuration is mechanically
(meta)stable. This implies that a region of size $N^*$:
\begin{equation} \label{Nstar1} N^* = \left(\frac{\gamma}{T
      s_c}\right)^{\frac{1}{1-x}},
\end{equation}
can always reconfigure. What happens physically is that the center of
the free energy distribution from Eq.~(\ref{Z5}) moves to the right
with $N$ according to $\gamma N^x$ because the mismatch typically
increases with the interface area. This alone would lead to a
depletion of states that are degenerate with the original state, which
is typically at $G_i = 0$.  Yet as $N$ increases, the free energy
distribution also {\em grows} in terms of the overall area, height,
and, importantly, width, as more states become available. For a
sufficiently large size $N^*$, the distribution is so broad that the
region is guaranteed to sample a state at $G_i = 0$ even though the
distribution center is shifted to the right by $\Gamma$. One may say
that in such a state, a negative fluctuation in the free energy
exactly compensates the mismatch penalty. For this to be typically
true, we must have
\begin{equation} \label{GammadG} \Gamma(N) = \delta G_i(N)
  \hspace{5mm} \text{ at } \hspace{5mm} N = N^*,
\end{equation}
where we have emphasized that both $\Gamma$ and $\delta G$ depend on
$N$.

Finally note that the physical extent $\xi$ of the reconfiguring
region:
\begin{equation} \label{xi} \left(\frac{\xi}{a} \right)^3 \equiv
  \frac{4 \pi}{3} \left(\frac{R^*}{a} \right)^3 \equiv N^*
\end{equation}
yields the volumetric cooperativity length for the reconfigurations.

\begin{figure}[t]
 \includegraphics[width=\figurewidth]{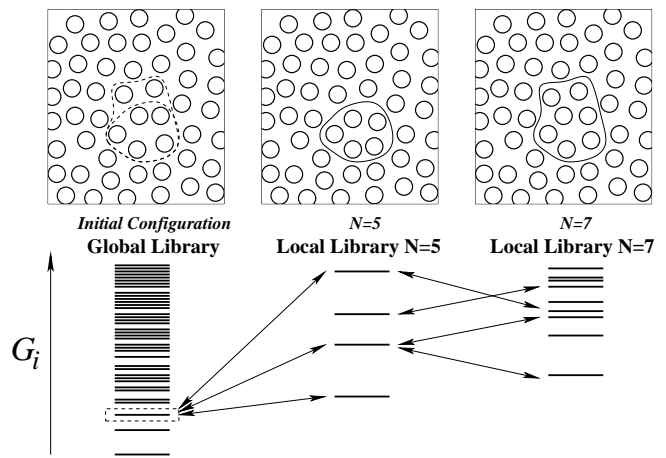}
 \caption{\label{library} Illustration of the library construction of
   aperiodic states.~\cite{LW_aging} On the left, we start out with
   some metastable structure. The density of the horizontal bars
   reflects the increase of the density of states (DoS)
   $e^{S_c(G_i)/k_B}$ with $G_i$. This DoS is distinct from the
   probability distribution in Eq.~(\ref{Z5}), which also includes the
   Boltzmann factor $e^{-\beta G_i}$.  In the center and right panels,
   5 and 7 particles have been moved. The density of states pertaining
   to the corresponding local regions are much lower than the global
   density of states.  In addition, the majority of the thus obtained
   configurations are higher in free energy than the original
   configuration, owing to the mismatch with the environment. As the
   region size grows, the distribution of free energies $G_i$ of
   individual structures is determined by a competition between a
   depletion due to the mismatch and an entropically-driven increase
   in the DoS. For a large enough $N = N^*$, there will be a
   configuration whose free energy $G_i$ is comparable to that of the
   original structure.}
\end{figure} 

The above notions can be discussed explicitly in terms of particle
movements, within the library construction of aperiodic
states,~\cite{LW_aging} graphically summarized in Fig.~\ref{library}.
We start out with the original state, which is some state from the
full set of states available to the system. We then draw a surface
encompassing a compact region containing $N$ particles and consider
all possible configurations of the particles inside while the
environment is static up to vibration. Most of the resulting
configurations are, of course, very high in energy because of steric
repulsion between the particles comprising the region itself and
between the particles on the opposite sides of the the boundary. Only
few configurations will contribute appreciably to the actual ensemble
of states, and even those are offset upwards, free energy-wise, by the
mismatch penalty. As the chosen region is made progressively bigger,
three things happen at the same time: (a) the mismatch penalty
typically increases as $\gamma N^x$; (b) the log-number of available
states first decreases but then begins to increase, asymptotically as
$\propto N$; and (c) the spectrum of states becomes broader, the width
going as $\propto \sqrt{N}$. Think of (a) as a density of states
$e^{S(G_i)/k_B}$ that moves up according to $\gamma N^x$ with the
region size (see Fig.~\ref{library}), thus leading to a ``depletion''
of the density of states at low $G_i$. Items (b) and (c), on the other
hand, mean the density of states increases in magnitude roughly as
$e^{s_c N -\Gamma - (G_i-\Gamma)^2/2 \delta G^2}$, at fixed $G_i$
($\delta G \propto \sqrt{N}$ and $\Gamma \propto N^x$). For a large
enough $N$, this growth of the density of states, at fixed $G_i$,
dominates the depletion due to the mismatch penalty and so the free
energy of the substituted configuration eventually stops growing and
begins to decrease with the region size $N$, after the latter reaches
a certain critical value $N^\ddagger$. Eventually, at size $N^*$, one
will typically find an available state that is mechanically stable.

The discussion of the statistical notions embodied in
Eqs.~(\ref{Gamma1})-(\ref{Nstar1}) in terms of particle motions helps
one to recognize that the full set of configurations in some range
$[0, N_\smax]$ can be sorted out into (overlapping) subsets according
to the following protocol: (a) within each subset, every region size
is represented at least once and (b) two configurations characterized
by sizes $N$ and $(N+1)$ differ by the motion of exactly one
particle. The subsets thus correspond to dynamically-connected paths,
along each of which particles join the reconfiguring region {\em one
  at a time}. Along each of the dynamically connected paths, one could
thus think of the reconfiguration as a {\em droplet} growth.  A
certain path will dominate the ensemble of the paths given a
particular final configuration.  This dominating path is the one that
maximizes the number of states $Z(N)$ from Eq.~(\ref{Z6}). This is
entirely analogous to the Second Law, whereby the equilibrium
configurations are those the maximize the density of states.

One may question whether the most likely bottle-neck configuration in
the set of all dynamically connected paths leading to the final state
at $N^*$ is, in fact, as likely as what is prescribed by
$Z(N^\ddagger)$ with $Z(N)$ from Eq.~(\ref{Z6}). The answer is yes
because sampling of all possible shapes and locations for a region of
size $N$ is implied in the summation in Eq.~(\ref{Z5}). By
backtracking individual dynamically-connected trajectories from $N =
N^*$ to $N = 0$, we can determine the precise reconfigured region that
produces the most likely bottle-neck configuration with probability
$Z(N^\ddagger)$.

Now, for region sizes in excess of $N^*$, the number $Z(N)$ of
available states exceeds one, implying that, for instance, {\em two}
distinct metastable configurations are available to a region of size
$N$ such that $Z(N) = 2$. According to the above discussion, the
trajectories leading to these two states are generally distinct,
though the probabilities of the respective bottle-neck configurations
and the corresponding critical sizes $N^\ddagger$ should be
comparable.

Structural reconfiguration can be equally well discussed not in a
microcanonical-like fashion, through the number of states $Z(N)$, but,
instead, in terms of the corresponding free energy $F(N) = - k_B T \ln
Z(N)$, as was originally done by Kirkpatrick, Thirumalai, and
Wolynes.~\cite{KTW} This yields the following activation profile for
the reconfiguration:
\begin{equation} \label{FN1} F(N) = \Gamma - T s_c N,
\end{equation}
Since we are considering all possible configurations for escape,
subject to the appropriate Boltzmann weight, this amounts to locally
replacing the original configuration encompassing $N$ particles by the
{\em equilibrated} liquid while the particles in the surrounding are
denied any motion other than vibration.  Upon the replacement, the
local {\em bulk} free energy is typically lowered by $\overline{G} -
\overline{G}_i = - T s_c N$, hence the driving term $- T s_c N$ in
Eq.~(\ref{FN1}).

The equilibrated liquid is a Boltzmann-weighted average of
alternative, metastable aperiodic structures that are mutually
distinct and are also generally distinct from the initial
configuration.  Another way of saying two structures are distinct is
that the particles belonging to the structures inside and outside do
not fit as snugly---at the interface between the structures---as they
do within the respective structures. This is quite analogous to the
mismatch between two distinct crystalline polymorphs in contact, such
as during a first order transition between the polymorphs. In contrast
with a polymorphic transition, the scaling of the mismatch penalty
with the area of the interface will turn out to be somewhat
complicated, whereby $x \ne (D-1)/D$.

Combining the free energy view with the notion of dynamically
connected trajectories, due to the library construction, we conclude
that the activation profile in Eq.~(\ref{FN1}) is also a {\em
  nucleation} profile.  Naturally, the bottle-neck configuration
corresponding to $N = N^\ddagger$ from Eq.~(\ref{Ncr1}) thus
corresponds to the critical nucleus size.  The corresponding barrier
is equal to
\begin{equation} \label{F1} F^\ddagger \equiv F(N^\ddagger) = \gamma
  \left(\frac{x\gamma}{T s_c } \right)^{\frac{x}{1-x}} (1-x).
\end{equation}
This directly shows that the escape rate from a specific aperiodic
state is indeed finite.

Yet there is more to the activation profile in Eq.~(\ref{FN1}). In
ordinary theories of nucleation, the nucleus continues to grow
indefinitely once it exceeds the critical size, unless it collides
with other growing nuclei, as it happens during crystallization, or,
for instance, when the supply of the contents for the minority phase
runs out, as it happens during fog. This essentially unrestricted
growth takes places because this way, the system can minimize its free
energy by fully converting to the minority phase. This view is
adequate when there are only two free energy minima to speak of and
the system converts between those two minima.

However in the presence of an exponentially large number of free
energy minima, we must go about the meaning of the $F(N)$ curve more
carefully. We must recognize that both the initial and {\em final}
state for the escape event are individual aperiodic states that are,
on average, equally likely. In fact, because we have chosen $G_i = 0$
as our free energy reference, $F(N)$ gives exactly the log-number
(times $-k_B T$) of thermally available states to the selected region.
As a result, that the free energy $F(N)$ reaches its initial value of
zero indicates that a mechanically metastable state is available to
escape to.  One is accustomed to situations in which the initial and
final state for a barrier-crossing event are {\em minima} of the free
energy, which does not seem to be the case in the above argument.
There is no paradox here, however. The quantity $F(N)$ is not the
actual free energy of the system. Instead, by construction, it is the
free energy under the constraint that the outside of the selected
compact region be not relaxing in the usual matter, but, instead, be
forced to be in a specific, metastable aperiodic minimum.  The
monotonic decrease of $F(N)$ at $N=N^*$ is trivial in that it simply
says the surrounding of the droplet will eventually proceed to
reconfigure again and again, as it should in equilibrium. As we
already emphasized, the state to which the initial configuration has
escaped is perfectly meta{\em stable}.

We now shift our attention to the {\em energy}. Suppose the liquid is
composed of particles that are not completely rigid, and so the
mismatch penalty has an energetic component. Furthermore, it is
instructive to suppose that the penalty is {\em mostly} energetic,
which is probably the case for covalently bonded substances such as
silica or the chalcogenides.~\cite{ZLMicro1} At a first glance, the
energy of the system appears to grow with each nucleation event, since
the driving force in Eq.~(\ref{FN1}) is exclusively entropic, at
equilibrium. Such unfettered energy growth is, of course, impossible
in equilibrium. On the contrary, the configurations before and after a
reconfiguration are typical and the energy must be conserved, on
average. The energy change following a transition must be within the
typical fluctuation range, which reflects the heat capacity $C_V$ at
constant volume and the bulk modulus $K$:~\cite{LLstat}
\begin{equation} \label{deltaE} \delta E = \{k_B C_v T^2 - V [T(\prtl
  p/\prtl T)_V -p]^2T/K \}^{1/2}.
\end{equation}
Note both $C_V$ and $V$ pertain to a single cooperative region. The
conservation of energy, on average, means that since one new interface
appears following an escape event, an equivalent of one interface must
have been {\em subsumed} during an event, as emphasized in
Ref.~\onlinecite{ZLMicro2}.

We thus conclude that the equilibrium concentration of the interface
configurations is given by $1/\xi^3$ with $\xi$ from Eq.~(\ref{xi}),
and so a glassy liquid is a {\em mosaic} of aperiodic
structures,~\cite{XW} each of which is characterized by a relatively
low free energy density, while the interfaces separating the mosaic
cells are relatively stressed regions characterized by excess free
energy density due to the mismatch between stabilized regions. This
stress pattern is not static, but relaxes at a steady pace so that a
region of size $\xi$ reconfigures once per time $\tau$, on average:
\begin{equation} \label{tauF1}
\tau = \tau_0 e^{F^\ddagger/k_B T},
\end{equation}
where the pre-exponent $\tau_0$ corresponds to the vibrational
relaxation time.

Note that the total free energy stored in the strained regions
corresponding to the domain walls is equal to $\Gamma (N/N^*) = T s_c
N$, i.e. the enthalpy difference between the liquid and the
corresponding crystal at the temperature in question, up to possible
differences in the vibrational entropy between the crystal and an
individual aperiodic structure.

According to Eqs.~(\ref{Ncr1})-(\ref{xi}), we need to evaluate the
exponent $x$ and the coefficient $\gamma$ for the mismatch penalty, to
estimate the escape rate and the cooperativity size for the activated
reconfigurations, to which we proceed next.

\section{Mismatch Penalty between Dissimilar Aperiodic Structures:
  Renormalization of the surface tension coefficient }
\label{mismatch}

Because the states on both sides of our interface are aperiodic, the
degree of mismatch is distributed.  Thus in some places the two
structures may fit quite well and so the scaling of the surface energy
term $\Gamma$ from Eq.~(\ref{FN1}) with the droplet size $N$ may be
weaker than the $N^{(D-1)/D}$ scaling expected for interfaces
separating periodic or spatially uniform phases.  The mechanism of
this partial lowering of the mismatch penalty is as follows: The
number of distinct aperiodic structures available to a sufficiently
large region, we remind, scales exponentially with the region
size. The free energies $G_i$ of individual structures from
Eq.~(\ref{gequil}) are {\em distributed}; they are equal on average
but differ by a finite amount for any specific pair of aperiodic
states.  Fluctuations of extensive quantities scale with $\sqrt{N}$ as
functions of size $N$.~\cite{LLstat} (The size $N$ at which the
$\sqrt{N}$ scaling sets in can be rather small in the absence of
long-range correlations, such as those typical of a critical point.)
Thus the free energy difference between the configurations outside and
inside scales as $\sqrt{N}$, for two regions of the same size $N$, and
could be of either sign.  Suppose now, for concreteness, that the
configuration on the outer side of the domain wall happens to be lower
in free energy than the adjacent region on the inside. Imagine
distorting the domain wall so as to replace a small portion of the
inside configuration by that from the outside.  It turns out the free
energy stabilization due to the replacement outweighs the
destabilization due to the now increased area of the interface, as we
shall see shortly.

Before we proceed with this analysis, it is instructive to discuss why
such surface renormalization and the consequent stabilization would
{\em not} take place during regular discontinuous transitions when one
phase characterized by a {\em single} free energy minimum nucleates
within another phase also characterized by a {\em single} free energy
minimum.  After all, both phases represent superpositions of
microstates whose energies are {\em also} distributed. Furthermore,
there seems to be a direct correspondence between, say, the regular
canonical ensemble and the situation described in Eq.~(\ref{gequil}).
Hereby, the free energies $G_i$ in Eq.~(\ref{gequil}) seem to
correspond to the energies of the microstates, while the
configurational entropy $S_c$ seems to correspond to the full entropy
in the canonical ensemble.  One difference between the situation in
Eq.~(\ref{gequil}) and the canonical ensemble is that in the latter,
transitions between the microstates within individual phases occur on
times much shorter than the observation time or mutual nucleation and
nucleus growth. As a result, the energies of the phases on the
opposite sides of the interface are always equal to their {\em
  average} values. In contrast, the distinct aperiodic states from
Eq.~(\ref{gequil}) are long-lived. In fact, the fastest way to
inter-convert between those states is via creation of the very
interface we are discussing! In the canonical ensemble analogy, this
would correspond to having {\em individual} microstates on the
opposite sides of the interface as opposed to ensembles resulting from
averaging over all microstates (with corresponding Boltzmann
weights). Conversely, the situation in Eq.~(\ref{gequil}) would be
analogous to the canonical ensemble only at sufficiently long times
that much exceed the nucleation time from Eq.~(\ref{tauF1}). Note that
at such long times, we have identical, equilibrated liquid on both
sides and so there is no surface tension in the first place.

Now, the situation where the system can reside in long-lived states
whose free energies are distributed in a Gaussian fashion can be
equivalently thought of as a perfectly ergodic, equilibrated system in
the presence of a {\em static}, externally-imposed random field whose
fluctuations scale in the Gaussian fashion.  In the absence of this
additional random field, the mismatch penalty between such two regular
phases would be perfectly uniform along an interface with spatially
uniform curvature. The simplest system one can think of, in which this
situation is realized, is the random field Ising model:
\begin{equation} \label{RFIM} \cH = - J \sum_{i<j} \sigma_i \sigma_j -
  \sum_i h_i \sigma_i , \hspace{10mm} \sigma_i = \pm 1,
\end{equation}
where $J > 0$ while the Zeeman splittings $h_i$'s are random,
Gaussianly distributed variables. In the Hamiltonian above, if one
were to impose a strictly flat interface between two domains with
spins up and down, the domain wall would distort some to optimize the
Zeeman energy. However, the amount of distortion is also subject to
the tension of the interface between the spin-up and spin-down
domains. The overall lowering of the free energy, due to the interface
distortion, corresponds to the optimal compromise between these two
competing factors.  Likewise, a smooth interface between two distinct
aperiodic states will distort to optimize with respect to local bulk
free energy, which is distributed. The energy compensation will scale,
again, as the square root of the variation of the volume swept by the
interface during the distortion.  The final shape of the interface
will be determined by the competition between this stabilization and
the cost of increasing the area of the interface.

The mapping between the random field Ising model and large scale
fluctuations of the interface between aperiodic liquid structures was
exploited by Kirkpatrick, Thirumalai, and Wolynes (KTW) \cite{KTW},
who used Villain's argument \cite{Villain} for the renormalization of
the surface in RFIM to deduce how the droplet interface tension scales
asymptotically with the droplet size.  The mapping relies crucially on
the condition that the undistorted interface must not be too thick, as
it would be near a critical point. This assumption turns out to be
correct since the width of the undistorted interface is on the order
of the molecular length $a$.~\cite{RL_sigma0}

Let us now consider a variation on the KTW-Villain argument concerning
the surface tension renormalization. This argument produces the sought
scaling relation for the mismatch penalty but some of its steps are
only accurate up to factors of order one, and so the latter will be
dropped in the calculation. All lengthscales will be expressed in
terms of the molecular length $a$, which simply sets the units of
length. Now, consider two dissimilar aperiodic states in contact, and
assume we have already coarse-grained over all length-scales less than
$r$, while explicitly forbidding interface fluctuations on greater
lengthscales. The interface is thus {\em taut}. Further, consider
spatial variations in the shape of the interface on lengthscales
limited to a narrow interval $[r, r(1+ \Delta)]$. The dimensionless
increment
\begin{equation} \label{increment}
  \Delta = d \ln r
\end{equation}
is the increment of the running argument for our real-space
coarse-graining transformation $r \to r(1+ \Delta)$.  (Ultimately, $r$
will be set at the droplet radius).  We may assume, without loss of
generality, that the mismatch penalty may be written in the following
form, in $D$ spatial dimensions:
\begin{equation} \label{Fsr} \Gamma = \sigma(r) r^{D-1},
\end{equation}
The quantity $\sigma(r)$ may be thought of as a renormalized surface
tension coefficient, where the amount of renormalization generally
depends on the wavelength, which is distributed in the (narrow) range
between $r$ and $r(1+\Delta)$. Our task is to determine under which
condition such renormalization takes place, if any.

To do this, let us deform the interface so as to create a bump of
(small) height $\zeta$ and lateral extent $r$.  Because the interface
is taut, the area will increase quadratically with $\zeta$. The
resulting increase in the interface area will incur a free energy cost
\begin{equation}
  \delta F_s \sim \sigma(r) r^{D-1} (\zeta/r)^2 \Delta,
\end{equation}
when $\zeta \ll r$.  It will turn out to be instructive to use a more
general form
\begin{equation} \delta F_s \sim \sigma(r) r^{D-1} (\zeta/r)^z \Delta.
\end{equation}
This generalized form is convenient because (a) rough interfaces may
exhibit a $z$ other than $2$ 
(b) the scaling of the interface tension with $r$ will turn out be
independent of $z$ at the end of the calculation. Thus the obtained
$\sigma$ vs. $r$ scaling can be argued to still apply even to
situations when $\zeta/r$ is not necessarily small. (Which it will not
be!)  Now, as already mentioned, one can always flip a region (of size
$N$) at the interface so a to lower its bulk free energy by $\sim h
\sqrt{N}$. The resulting bulk free energy gain is thus:
\begin{equation} \label{Fb} \delta F_b \sim - h \sqrt{N} \Delta \sim -
  h (r^{D-1} \zeta)^{1/2} \Delta,
\end{equation}
where the constant $h$ is straightforward to estimate in light of our
earlier discussion that the bulk stabilization above is the result of
fluctuations of the Gibbs free energy.  Thus,
\begin{equation} \label{h2} h \sim \delta G_i /\sqrt{N},
\end{equation}
with $\delta G_i$ from Eq.~(\ref{dGi}).  Properly, we should have
written $h^2 = 2 (\delta G_i)^2/N$ in Eq.~(\ref{h2}) because the bulk
free energy stabilization is a {\em difference} between two random
Gaussian variables, whose distribution widths are $\delta G_i$ each,
but we have agreed to drop factors of order one in the derivation, see
also below.

Finally, one should not fail to identify the stabilization in
Eq.~(\ref{Fb}) with the Bouchaud-Biroli free energy stabilization of a
compact region due to matching with its surrounding.

Next we find the value of $\zeta$ that minimizes the total free energy
stabilization: $\prtl (\delta F_s + \delta F_b)/\prtl \zeta = 0$,
which yields:
\begin{equation} \label{zeta} \zeta \sim (h/\sigma)^{2/(2z-1)}
  r^{(2z-D+1)/(2z-1)}.
\end{equation}
The resulting energy gain per unit area,
\begin{align} \label{minF}
  \min_\zeta\{& \delta F_s + \delta F_b\}/r^{D-1} \sim \nonumber \\
  &\sim - (h^z/\sigma^{1/2})^{1/(z-1/2)} r^{-z(D-2)/(2z-1)} \Delta,
\end{align}
thus represents the renormalization $\delta \sigma(r)$ of the
$r$-dependent ``surface tension coefficient'' that resulted from
integrating out degrees of freedom in the $k$-vector range between
$1/r$ and $1/(1+\Delta) r$.

The energy gain per unit area from Eq.~(\ref{minF}), due to the
real-space renormalization in the wavelength range $[r, r(1+\Delta)]$,
can be viewed as an iterative relation, by Eq.~(\ref{increment}):
\begin{equation} \label{dsdr}
  d \sigma \sim - (h^z/\sigma^{1/2})^{1/(z-1/2)} r^{-z(D-2)/(2z-1)} 
  d \ln r.
\end{equation}
A quick inspection of this differential equation shows that the
surface tension coefficient {\em decreases} with $r$. To determine the
actual $r$-dependence of $\sigma$, we must decide on the boundary
condition $\sigma(r=\infty) \equiv \sigma_\infty$. Suppose for a
moment that $\sigma_\infty > 0$, which implies that at sufficiently
large distances, the interface width tends to some {\em finite} value
$l_\infty$, however large. This is because the surface tension
coefficient of a {\em flat} interface between non-matching
equilibrated structures goes as $g^\ddagger l_\infty$, where
$g^\ddagger$ is the local free energy density excess at the
interface.~\cite{RowlinsonWidom, Bray} Since the free energy excess
per unit volume $g^\ddagger$ tends to a finite value
$g^\ddagger_\infty > 0$ in the limit of a flat interface, so should
$l_\infty$, if $\sigma_\infty$ is finite.  Now, the surface tension
coefficient is thus given by the expression
\begin{equation*} \sigma^{2z/(z-1)}(r) = h^{2z/(2z-1)}
  r^{-z(D-2)/(2z-1)} + \sigma^{2z/(z-1)}_\infty.
\end{equation*}
Inserting the above formula in expression (\ref{zeta}) yields:
\begin{equation} \label{zeta1} \zeta \sim \frac{r^{(2z-D+1)/(2z-1)}}
  {[r^{-z(D-2)/(2z-1)}+ (\sigma_\infty/h)^{2z/(2z-1)} ]^{1/z}}
\end{equation}
The above formula indicates that although incremental changes in the
interface curvature following the renormalization are small, the
compound increase in the interface thickness---due to the curvature
changes in the broad wavelength range spanned by the coarse-graining
procedure---is not necessarily so.

Eq.~(\ref{zeta1}) indicates that there are two internally-consistent
options regarding the value of the surface tension coefficient
$\sigma_\infty$. In the conventional case of zero random field, $h =
0$, $\sigma_\infty$ is finite while $\zeta = 0$, and so no
renormalization takes place while the interface width tends to a
steady value $l_\infty$ at diverging droplet radii. If, on the other
hand, the random field is present, the only remaining option is
$\sigma_\infty = 0$. Indeed, by Eq.~(\ref{zeta1}), the interface width
$\zeta$ diverges as $r \to \infty$, when $h > 0$, implying the
supposition of a finite $\sigma_\infty$ and, hence, finite $l_\infty$
was internally inconsistent.  For this argument to be valid, the
renormalized interface width $\zeta$ should exceed the width $l$ of
the original, smooth interface. Condition $\zeta > l$ and
Eq.~(\ref{zeta1}), combined with $\sigma_\infty = 0$, yield
\begin{equation} r > l,
\end{equation}
which happens to coincide with the criterion of validity of the thin
interface approximation. Note that in their analysis of barrier
softening effects near the crossover, Lubchenko and
Wolynes~\cite{LW_soft} self-consistently arrived at a similar
criterion, viz., $r > a$, for when interface tension renormalization
would take place.

It follows that an arbitrarily weak, but finite random field $h$ makes
the interface with a sufficiently low curvature unstable with respect
to distortion and lowering of the effective surface tension:
\begin{equation} \label{sr} \sigma(r) \sim \frac{h}{a^{D-1}}
  (r/a)^{-(D-2)/2},
\end{equation}
independent of $z$, apart from a proportionality constant of order
one, giving us confidence in the result even when the undulation size
$\zeta$ is not very small. Notice we have restored the units of length
for clarity. 

Eq.~(\ref{sr}) yields that the renormalized mismatch energy $\Gamma$
from Eq.~(\ref{Fsr}) scales with the droplet size $N$ is a way that is
independent of the space dimensionality, namely $\sqrt{N}$:
\begin{equation} \label{Gr} \Gamma \sim \sigma(r) r^{D-1} \sim h
  (r/a)^{D/2} \sim h \sqrt{N},
\end{equation}
which is ultimately the consequence of the Gaussian distribution of
the free energy.  Because of the lack of a fixed length scale in the
problem---other than the trivial molecular size $a$, which sets the
units---it should not be surprising that the interface width $\zeta$
scales with the radius $r$ itself:
\begin{equation}
  \zeta \sim r,
\end{equation}
again independent of $z$. The numerical constant in the above equation
is of order one, as is easily checked, and so $\zeta/r$ is not small
generally.

The large effective interface width can be thought of as a result of
the distortion of the original thin interface where the extent of the
distortion is not determined by a fixed length, but the curvature of
the interface itself. In other words, this interface is a {\em
  fractal} object. Because of this fractality, the structure at the
interface is not possible to characterize as either of the aperiodic
structures on the opposite sides of the original smooth interface
before the renormalization. We could thus informally think of this
fractal interface as the original thin interface {\em
  wetted}~\cite{XW} by other structures that interpolate, in an
optimal way, between the two original aperiodic structures. While we
are not aware of direct molecular studies with regard to the
fractality of cooperative regions in non-polymeric liquids, such
studies of polymer melts do suggest the mobile regions are fractal in
character.\cite{Starr2014}

According to Eq.~(\ref{Gr}), the scaling exponent $x$ is equal to
$1/2$. Thus the matching condition in Eq.~(\ref{GammadG}) is valid at
all values of $N$, while
\begin{equation} \label{Gamma} \Gamma = \gamma \sqrt{N}, 
\end{equation}
where $\gamma = \delta G_i/\sqrt{N} = \text{const}$, thus yielding:
\begin{eqnarray} \label{gammaFull} \gamma &=& \left\{ \left[ K - T
      \left(\frac{\prtl S_c}{\prtl V} \right)_T \right]^2 \frac{k_B
      T}{K \bar{\rho}} \right. \nonumber \\ &+& \left. [K \alpha_t + (
    \Delta \tilde{c}_v - \tilde{s}_\svibr) ]^2 \frac{k_B
      T^2}{\bar{\rho} \tilde{c}_v} \right\}^{1/2}.
\end{eqnarray}

In retrospect, the square-root scaling in Eq.~(\ref{Gamma}) is
natural: In view of Eq.~(\ref{Gamma1}), the $(G_i-\Gamma)^2/2 \delta
G_i^2$ term under the second exponential in Eq.~(\ref{Z5}) scales
asymptotically with $N$ according to $N^{2x-1}$, see also
Ref.~\onlinecite{capillary}.  For any $x$ other than $1/2$, this would
result in an anomalous scaling~\cite{Goldenfeld} of the density of
states with the system size that would be hard to rationalize given
the apparent lack of criticality in actual liquids between the glass
transition and fusion temperatures.

Note that the result in Eq.~(\ref{Gamma})--(\ref{gammaFull}) is only
approximate because it does not include finite size corrections.
Apart from these corrections, we obtain for the nucleation barrier
from Eq.~(\ref{F1}):
\begin{equation} \label{F2} F^\ddagger = \frac{\gamma^2}{4T s_c }.
\end{equation}

Next we estimate $\gamma$. As a rule of thumb, the bulk modulus is
about $(10^1 - 10^2) k_B T/\bar{\rho}$ for liquids and $10^2 k_B
T/\bar{\rho}$ for solids near the melting temperature
$T_m$~\cite{Bilgram} (consistent with the Lindemann criterion of
melting~\cite{Lindemann, L_Lindemann}).  The rate of change of the
configurational entropy with volume is not known but can be crudely
estimated based on the observation that upon freezing, the hard sphere
liquid loses $\approx 1.2 k_B$ worth of entropy per particle while its
volume reduces by about $10$\%.~\cite{HooverRee1968, Hansen} Assuming
our liquid will run out of configurational entropy at about the same
rate---though gradually---we obtain $(\prtl S/\prtl V)_T \sim 10^1
k_B$.  Further, $\tilde{s}$ is about $10^0 k_B/a^3$. The dimensionless
expansivity $\alpha_t T$ is generically $10^{-1}$, although could be
much smaller for strong substances, see Fig.~12 of
Ref.~\onlinecite{RL_LJ}. $\Delta \tilde{c}_v$ and $\tilde{s}_c$ are both
$\sim 10^0 k_B/a^3$. As a result, we conclude that the volume
contribution to the free energy fluctuation in Eq.~(\ref{dGi}) exceeds
the temperature contribution by about two orders of magnitude or so,
thus yielding:
\begin{equation} \label{gamma1} \gamma^2 \approx \left[ K - T
    \left(\frac{\prtl S_c}{\prtl V} \right)_T \right]^2 \frac{k_B T}{K
    \bar{\rho}},
\end{equation}
since $\bar{\rho} \equiv a^{-3}$.

We thus obtain for the activation barrier:
\begin{equation} \label{FKsc} F^\ddagger \approx \left[ K - T
    \left(\frac{\prtl S_c}{\prtl V} \right)_T \right]^2 \frac{k_B}{4 K
    \tilde{s}_c},
\end{equation}
where $\tilde{s}_c$ is the configurational entropy per unit volume.

According to the above estimate of the $(\prtl S_c/\prtl V)_T$ term,
it is likely that at least for rigid and weakly attractive particles,
the second term in the square brackets is an order of magnitude
smaller than the first term.  We thus expect the following, simple
expression for the surface tension coefficient $\gamma$ to be of
comparable accuracy to Eq.~(\ref{gamma1}):
\begin{equation} \label{gamma2} \gamma \approx \left( \frac{K k_B T}{
      \bar{\rho}} \right)^{1/2} \equiv \sqrt{K a^3 \, k_B T}.
\end{equation}
It is interesting that the coefficient $\gamma$ above, which reflects
coupling of structural fluctuations to its environment, has exactly
the same form as the coupling between the structural reconfigurations
corresponding to the two-level systems (TLS) in cryogenic glasses and
the phonons.~\cite{LW, LW_RMP} Lubchenko and Wolynes have argued the
TLS correspond to the two lowest energy levels of the local degrees of
freedom that correspond to the low-barrier subset of the activated
reconfigurations near the glass transition temperature.~\cite{LW,
  LW_RMP}

The simplified form in Eq.~(\ref{gamma2}) implies for the nucleation
barrier:
\begin{equation} \label{FKsc1}
  F^\ddagger \simeq \frac{K}{4 ( \tilde{s}_c/k_B) }.
\end{equation}
This result can be compared with the earlier expression for the
activation barrier derived by Xia and Wolynes,~\cite{XW} viz.,
$F^\ddagger = 32. k_BT/(s_c/k_B)$, in which the configurational
entropy is per bead. In contrast, in expression (\ref{FKsc1}) the
configurational entropy is per unit {\em volume}.

Given that the temperature dependence of the bulk modulus is usually
rather weak, Eq.~(\ref{FKsc1}) approximately yields the venerable
Adam-Gibbs functional relation.~\cite{AdamGibbs} To avoid confusion we
emphasize that Eq.~(\ref{FKsc1}) applies {\em below} the
crossover. The Adam-Gibbs relation has also been used to fit
temperature dependences of relaxation data {\em above} the crossover,
but with coefficients that are generally different from those
determined in low temperature fits. The presence of a crossover
between distinct high-$T$ and low-$T$ Adam-Gibbs behaviors was brought
home elegantly by the numerical analysis of Stickel et
al.\cite{Stickel}

Finally, the expression for the cooperativity size $\xi \equiv a
(N^*)^{1/3} = (\gamma/T s_c)^{2/3}$ corresponding to the approximation
in Eq.~(\ref{FKsc1}) reads:
\begin{equation} \label{xiKsc} \xi \simeq \left[ \frac{K}{k_B T
      (\tilde{s}_c/k_B)^2 } \right]^{1/3}
\end{equation}
Note that neither of the expressions (\ref{FKsc}), (\ref{FKsc1}), and
(\ref{xiKsc}) depends explicitly on the molecular length scale $a
\equiv \bar{\rho}^{-1/3}$.

\section{Comparison with Experiment and Earlier Approximations}
\label{tests}

Because the quantity $(\prtl S_C/\prtl V)_T$ is not presently
available, we analyze the more approximate expression from
Eq.~(\ref{FKsc1}). As the primary test, we compare the present
predictions for the barrier with experiment and two earlier
predictions, in relatively extended temperature ranges.  These earlier
predictions~\cite{RWLbarrier} utilize the values of the surface
tension coefficient $\sigma_0$ from Eq.~(\ref{FRsc})---or $\gamma$
from Eq.~(\ref{F2})---as determined by Xia and Wolynes~\cite{XW} (XW)
and Rabochiy and Lubchenko~\cite{RL_sigma0} (RL) respectively. We
include in the present analysis all eight substances considered in
Ref.~\onlinecite{RWLbarrier} except toluene, for which bulk modulus data are
not available. The results for the remaining seven substances are
shown in Fig.~\ref{Cal}. Note that we have estimated the values of the
bulk modulus based on the speeds of longitudinal and transverse sound,
as determined by Brillouin scattering.  The latter procedure yields
the adiabatic value of the bulk modulus which somewhat exceeds its
isothermal value. It is the isothermal modulus that enters the present
estimates.  As in Ref.~\onlinecite{RWLbarrier}, we have chosen to plot the
barriers within the dynamical range representative of actual liquids,
viz. $\ln(\tau/\tau_0) \le 35.7$, which corresponds to a glass
transition on 1 hr time scale and $\tau_0 = 1$~ps. This way, the error
of the approximation exhibits itself through an error in the
temperature corresponding to a particular value of the relaxation
time.

We observe that the present predictions compare reasonably with both
experiment and the XW and RL approximations.  Although the barrier
values predicted with the help of all three approximations are in
general agreement with expectation, their deviation from measured
values is quite significant for some of the substances. The reader is
referred to our earlier detailed discussion\cite{RL_sigma0,
  RWLbarrier} on the possible sources of error for the elastic
constants and the configurational entropy.  Also there, the published
sources of the experimental data can be found. Here we only note that
determination of the elastic constant is affected by a large
dissipative component to the elastic response. On the other hand, the
estimations of the configurational entropy are subject to uncertainty
in the vibrational entropy difference between crystal and glass and
the determination of the (putative) Kauzmann temperature by
extrapolation of thermodynamic or kinetic data beyond the glass
transition temperature.

\begin{figure}[t]
\centering
\includegraphics[width = \columnwidth]{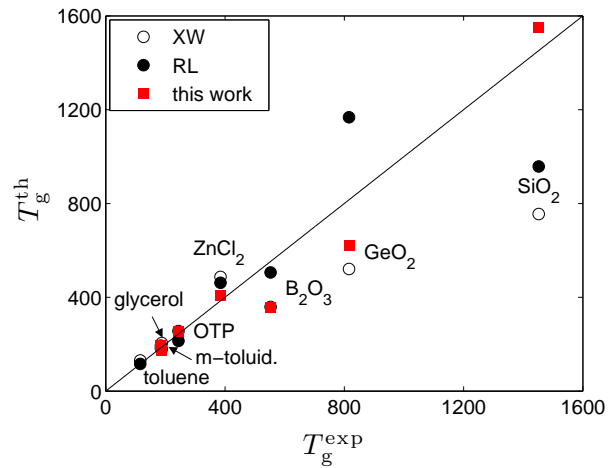}
\caption{\label{Tg} Theoretically predicted glass transition
  temperatures $T_g$, calculated using expression (\ref{FKsc1}),
  alongside the predictions~\cite{RWLbarrier} utilizing the
  XW~\cite{XW} and RL~\cite{RL_sigma0} approximations for the mismatch
  penalty, plotted against experimental values. The
  theoretically-determined temperatures are based on the barriers from
  Fig.~\ref{Cal} and rely on static input data, while the experimental
  values are kinetic quantities obtained from calorimetry.}
\end{figure}

The results from Fig.~\ref{Cal} can be partially summarized by
graphing the predicted values for the glass transition temperature for
all three approximations against their experimental values, see
Fig.~\ref{Tg}. Clearly, theoretical predictions for the glass
transition correlate very well with experiment.

Both for the XW and RL calculations, the formulas for the barriers
explicitly contain the molecular size $a$. This molecular size is
associated with the size of the rigid molecular unit, which we have
called the ``bead.''  The bead size is well-defined for relatively
rigid, weakly attractive particles.~\cite{LW_soft} In covalently
bonded liquids, on the other hand, it can be also understood as the
ultra-violet cut-off of the theory.~\cite{BL_6Spin} In any case, a
reasonable estimate for $a$ can be obtained by calibrating the fusion
entropy of the substance by that of a Lennard-Jones
liquid.~\cite{LW_soft, StevensonW} We have called this way to count
beads ``calorimetric.'' Because the uncertainty in the bead size $a$
is potentially a significant contributor to the deviation of the XW
and RL predictions from experiment, it is worthwhile to vary it and
see whether the deviation can be minimized.  This has been done in
Ref.~\onlinecite{RWLbarrier} so as to bring the activation exponent
$F^\ddagger(T_g)/k_B T_g$ to a fixed value of 35.7 and results in
chemically-reasonable values of $a$, except for OTP.~\cite{RL_sigma0}
The barrier values corresponding to the so renormalized bead size are
shown in Fig.~\ref{SC} for the seven substances in question.

The present predictions, Eqs.~(\ref{FKsc}) and (\ref{FKsc1}), do not
depend on the bead size and so the discrepancy with experiment is
entirely due to the error in the approximation and in the experimental
values of the configurational entropy and the elastic constants.  In
Fig.~\ref{SC}, we show $F^\ddagger(T)/k_B T$, with $F^\ddagger$ from
Eq.~(\ref{FKsc1}), multiplied by a constant so as to bring the value
of $F^\ddagger(T_g)/k_B T_g$ to 35.7. We observe that the temperature
dependence of the so adjusted barrier is quite similar to experiment,
perhaps more so than either of the XW or RL approximation. (We remind
that the $T$-dependence of the RL-based barrier values cannot be fully
judged based on Figs.~\ref{Cal} and \ref{SC} because temperature
dependent data for the structure factor are not currently available;
as a result the values measured at $T_g$ were used.)  We note that the
experimental value for the $F^\ddagger(T_g)/T_g$ ratio is subject to
uncertainty in the prefactor $\tau_0$ and the cooling rate; the ratio
will thus generally differ from 35.7. This, however, is not much of an
issue with regard to comparing the {\em slopes} of the curves.

Finally we note that the cooperativity length $\xi$ can be
straightforwardly computed using Eq.~((\ref{xiKsc})). The XW and RL
based values of $\xi$ are very similar to each other and are close to
the experimentally deduced ones in the first place. In view of the
similarity of the values of the mismatch penalty computed with the
present approximation with those earlier approximations, the
cooperativity length (\ref{xiKsc}) is automatically numerically close
to those computed in Ref.~\onlinecite{RWLbarrier}.

\section{Discussion and Conclusions}
\label{conclusions}

This work has revisited the mechanism of the activated transport in
glassy liquids. We have highlighted a key notion that, we hope, will
clarify the entropic droplet scenario of the activated transport, due
to Wolynes and coworkers. The mutual conversion between metastable
(aperiodic) configurations shows an important distinction from mutual
conversion between two {\em equilibrated} phases, such as during a
conventional first order transition.  In the latter case, the
distribution of the energies of the microstates does not significantly
affect the nucleation rate even if the mismatch penalty between the
two phases strongly depends on the precise identity of the microstates
on the opposite sides of the interface. This is because the
microstates within the individual phases interconvert on times scales
much shorter than the duration of the nucleation event. As a result,
the mismatch penalty between two equilibrium phases is spatially
uniform along the interface (of constant curvature) and does not
deviate significantly from a certain average value. In contrast,
interconversion itself between distinct aperiodic structures is
mediated by creation of an interface between those structures. And so
the mismatch penalty will strongly depend on the precise identity of
the two states separated by the interface; this penalty can be lowered
by distorting the interface to give locally a bit more room to {\em
  that} state of the two which is lower in free energy, while not
moving the interface as a whole.

The magnitude of the mismatch penalty is essentially equal to the
magnitude of local free energy fluctuations. In turn, this magnitude
turns out to be straightforwardly related to thermodynamic properties
of the liquid, see Eqs.~(\ref{dGi}) and (\ref{gammaFull}), and is
largely determined by the bulk modulus of the liquid. In its most
approximate form, the reconfiguration barrier is given by the simplest
expression of units energy one could write down, Eq.~(\ref{FKsc1}),
using the configurational entropy per unit volume and the bulk
modulus; note the latter has dimensions energy density. The inverse
linear scaling of the barrier with the configurational entropy is
special in that it is the only scaling in which the barrier does not
explicitly depend on temperature. In turn, this special scaling is a
direct consequence of the $\sqrt{N}$ dependence of the mismatch
penalty.  Interestingly, the barrier does not explicitly depend on the
molecular length $a$.  To appreciate this, we note that the expression
for the cooperativity length $\xi$ does not explicitly contain $a$
either. This is consistent with the mismatch penalty originating from
ordinary structural fluctuations which have Gaussian statistics and
are scale free. Such fluctuations would not lead to a static
heterogeneity, consistent with the results of an independent argument
by Cammarota and Biroli.~\cite{0295-5075-98-3-36005} In fact, one may
think of the mismatch itself as a stress pattern corresponding to
frozen-in structural fluctuations or frozen-in {\em stress},
consistent with the elasticity-based picture of glassy liquids of
Bevzenko and Lubchenko.~\cite{BL_6Spin}

Although entropic droplet nucleation and mutual nucleation of two
coexisting {\em thermodynamic} phases are similar in some respects,
they also exhibit important differences, including distinct scalings
of the mismatch penalty with the droplet size. In addition, during
regular first order transitions, the two phases separated by the
interface are themselves minima of the free energy of the system. The
interfacial configurations may be thought of as corresponding to the
barrier that separates the two minima in the bulk term of the
Landau-Ginzburg functional that describe the two phases. As a result,
the physical space is either wholly occupied by one of the phases or
the two phases both occupy macroscopic regions, if they are in near
equilibrium. The interfacial configurations thus occupy a negligible
portion of the sample.  In contrast, the contents of an entropic
droplet by themselves do {\em not} correspond to a minimum of the free
energy. The metastable configuration in the {\em beginning} of an
activated event ($N=0$) does correspond to a free energy minimum, and
so does the configuration at the {\em end} of an activated event, upon
which a compact region of size $N = N^*$ has been replaced by an
alternative structure. The border between the reconfigured region and
its environment is generally characterized by a higher free energy
density. The original configuration {\em also} contained similarly
strained regions that had resulted from preceding reconfigurations at
the locale in question; the energy is thus conserved, on average, as a
result of activated events. One may think of a glassy liquid as a
mosaic of relatively stabilized regions separated by relatively
strained regions. These strained regions can be thought of as
transiently frozen-in stress, see Ref.~\onlinecite{BL_6Spin} and,
also, very recent work by the same authors,~\cite{BLelast} in which
alternative procedures of defining the built-in stress are carefully
discussed.

The present discussion gives a relatively simple perspective on one of
the most intriguing aspects of the structural glass transition: In
spin glasses, as in the Sherrington-Kirkpatrick
model,~\cite{SpinGlassBeyond} the disorder is quenched. In contrast,
the molecular interactions are translationally invariant and so the
disorder in liquids is {\em self-generated}. What is the mechanism of
this self-generation? The classical density functional theory has
shown conclusively that aperiodic structures become metastable at
sufficiently high densities.\cite{dens_F1, dens_F2, BausColot, Lowen,
  RL_LJ} Within the replica methodologies,~\cite{PhysRevLett.75.2847,
  PhysRevLett.82.747} one can arrive at the disorder by considering,
for instance, a random field generated by a generic boundary for a
compact liquid region and then determining this field
self-consistently. Within the present discussion, the random field is
indeed created by an environment composed of aperiodic structures. The
field is effectively static---not unlike the quenched disorder in spin
models---but only on the time scale of the activated reconfigurations
and thus {\em transiently breaks} the translational symmetry.

Because the environment is static, the free energy profile $F(R)$ from
Eq.~(\ref{FR})---or $F(N)$ from Eq.~(\ref{FN1})---is not the actual
free energy of the system. Instead, it gives the log-number (times
$-k_B T$) of the states thermally available to a chosen compact region
under the constraint that its environment be static up to
vibration. The value of $F(N)_{N=N^*} = 0$ is thus special in that it
signifies the cooperativity size $N^*$ for the reconfigurations
because an alternative, mechanically metastable state is available at
$N = N^*$. The downhill decrease of $F(N)$ at $N=N^*$ does not mean
that an individual nucleation event will proceed indefinitely past
$N^*$ but, instead, that the environment itself will eventually
proceed to reconfigure. Alternatively, one may think of this downhill
decrease as reflecting that more than one aperiodic state with a
comparable energy---within the fluctuation range in
Eq.~(\ref{deltaE})---is available to a region of size $N > N^*$. The
identity of the state to which the liquid will reconfigure is subject
to the precise barrier separating that state from the initial
configuration. Note that like the bulk free energy of the individual
aperiodic states, this barrier is also distributed. The width of this
distribution is determined by the configurational heat capacity, which
is ultimately responsible for the correlation between the heat
capacity jump at the glass transition and the degree of
non-exponentiality of liquid relaxations, as established by Xia and
Wolynes.~\cite{XWbeta} The barrier distribution is also instrumental
in the violation of the Stokes-Einstein relation and decoupling
between various processes in supercooled liquids.~\cite{XWhydro,
  Lionic}

We finish by pointing out several convenient relationships between
material constants and various characteristics of the glass
transition. For instance, the configurational entropy can be estimated
according to the expression:
\begin{equation} \tilde{s}_c = \frac{K}{4T \ln(\tau/\tau_0)}.
\end{equation}
In view of the near universality of $\ln(\tau/\tau_0)$ at the glass
transition, where it is close to 35 or so, we obtain that
\begin{equation} \tilde{s}_c(T_g) \simeq 10^{-2} \frac{K(T_g)}{T_g}
  \sim \frac{k_B}{a^3}.
\end{equation}

Likewise, the cooperativity size can be written as
\begin{equation} \label{xiK} \xi^3 \equiv N^* a^3 = [4
  \ln(\tau/\tau_0)]^2 \frac{k_B T}{K},
\end{equation}
Again note $\xi$ does not explicitly depend on the molecular length
$a$. According to Eq.~(\ref{xiK}), the cooperative size near the glass
transition is on the order of a few hundred rigid molecular units:
\begin{equation}
N^*(T_g) \sim 10^2.
\end{equation}

Lubchenko and Wolynes~\cite{LW, LW_RMP} have argued that the density
of states of the two-level systems in cryogenic glasses is
approximately equal to $n_\text{TLS} \simeq 1/k_B T_g \xi^3(T_g)$. By
Eq.~(\ref{xiK}), this yields:
\begin{equation} \label{nTLS}
  n_\text{TLS} \simeq \frac{K(T_g)}{[4 k_B T_g \ln(\tau(T_g)/\tau_0)]^2 },
\end{equation}
where $K(T_g)$ is the isothermal bulk modulus just above the glass
transition.

Another interesting quantity is the Boson Peak frequency, which has
been estimated by Lubchenko and Wolynes~\cite{LW_BP} to be about
$\omega_\text{BP} \simeq (a/\xi) \omega_D$, where $\omega_D$ is the
Debye frequency. This prediction is consistent with
experiment.~\cite{PhysRevE.83.061508} Using $\omega_D \sim c_s/a$, we
get $\omega_\text{BP} \sim c_s/\xi$, where $c_s$ is the speed of
sound. Further noting that $c_s \simeq (K/\rho_M)^{1/2}$, where
$\rho_M$ is the mass density, and using Eq.~(\ref{xiK}), we obtain a
scaling relation
\begin{equation} \label{omegaBP} \omega_\text{BP} \sim
  \frac{K^{5/6}(T_g)}{(k_B T_g)^{1/3} \rho_M^{1/2} [4
    \ln(\tau(T_g)/\tau_0)]^{2/3}}.
\end{equation}

Lastly, consider the quantity $(\xi/a)^3$ that has been
argued~\cite{LW} to determine the ratio of the phonon mean-free path
$l_\text{mfp}$ to the the thermal phonon wavelength $\lambda$ in
cryogenic glasses. Within the XW approximation, the $(\xi/a)^3$ is
independent of the bead size $a$ and is universal at $T_g$. In the
present approximation, this ratio does explicitly depend on $a$ and
also on $K$:
\begin{equation} \label{xia3K} \frac{l_\text{mfp}}{\lambda} \simeq
  (\xi/a)^3 = [4 \ln(\tau/\tau_0)]^2 \frac{k_B T_g}{K(T_g) a^3}.
\end{equation}

In practice, the ratio of the bulk modulus to the glass transition
temperature does not vary all that much between different substances
because, on the one hand, the $K/T_m$ is nearly universal, owing to
the Lindemann criterion, while the $T_m/T_g$ ratio is often
numerically near 1.5, even though it does seem to vary between 1.2 and
1.6.  As a result, the cooperativity length $\xi$ will exhibit a great
deal of universality; the easiest way to vary it in experiment may be
to play with the cooling rate at the glass transition.  One the other
hand, the quantities (\ref{nTLS}) and (\ref{omegaBP}) scale with
energy and are immune to this potential limitation.

{\bf Acknowledgments:} We are indebted to Peter G. Wolynes for many
inspiring conversations. We thank Francesco Zamponi and Simone
Capaccioli for sharing fits of the experimental data for the
configurational entropy and relaxation times; and also Philip
S. Salmon for making available $S(k)$ data for several substances.  We
gratefully acknowledge the support by the NSF Grant CHE-0956127, the
Welch Foundation Grant No. E-1765, and the Alfred P. Sloan Research
Fellowship.

\appendix

\section{Fluctuations of the Gibbs free energy}

In the Appendix, we will drop the bars over the expectation values of
thermodynamical variables, for typographical convenience. In the
standard fashion,~\cite{LLstat} it is convenient to present
fluctuations of the Gibbs free energy in terms of fluctuations $\Delta
V$ and $\Delta T$ of volume and temperature respectively:
\begin{equation} \label{DG1} \Delta G = \left(\frac{\prtl G}{\prtl V}
  \right)_T \Delta V + \left(\frac{\prtl G}{\prtl T} \right)_V \Delta
  T,
\end{equation}
which yields
\begin{equation} \label{DG2} \la (\Delta G)^2 \ra = \left(\frac{\prtl
      G}{\prtl V} \right)^2_T \la (\Delta V)^2 \ra + \left(\frac{\prtl
      G}{\prtl T} \right)^2_V \la (\Delta T)^2 \ra,
\end{equation}
since $\la \Delta V \Delta T \ra = 0$.~\cite{LLstat} Further using
\begin{equation} \label{incrG} dG = - SdT + V dp
\end{equation}
to compute the derivatives in Eq.~(\ref{DG1}), the
equality~\cite{LLstat}
\begin{equation} \label{DV} \la (\Delta V)^2 \ra = - k_B T (\prtl
  V/\prtl p)_T \equiv k_B T V/K,
\end{equation}
and the equalities $\la (\Delta T)^2 \ra = k_B T^2/C_v$, $(\prtl
p/\prtl T)_V = - (\prtl p/\prtl V)_T (\prtl V/\prtl T)_p$ one
straightforwardly obtains Eq.~(\ref{dG}) of the main text.

Analogously for the Gibbs free energy of individual aperiodic states,
\begin{equation} \label{DGi} \Delta G_i = \left(\frac{\prtl G_i}{\prtl
      V} \right)_T \Delta V + \left(\frac{\prtl G_i}{\prtl T}
  \right)_V \Delta T.
\end{equation}
Eqs.~(\ref{gequil}) and (\ref{incrG}) yield
\begin{equation} \label{incrGi} dG_i = - S_\svibr dT + TdS_c + Vdp.
\end{equation}
Repeating the steps above, one arrives at Eq.~(\ref{dGi}) of the main
text. One should note that Eq.~(\ref{DV}) does not apply to solids, in
which the shear modulus $\mu > 0$. In the latter case, volume
fluctuations are of lower magnitude and are given by the expression
$\la (\Delta V)^2 \ra = k_B T V/(K + 4\mu/3)$.~\cite{RL_sigma0} Note
that here we are dealing with an {\em equilibrated} liquid, for which
$\mu = 0$. Alternatively said, the full distribution of the free
energies $G_i$ of individual aperiodic states accounts for not only
vibrations within individual structures but for the {\em full} variety
of the structures. Sampling over this variety corresponds to
equilibration of the liquid, which corresponds with $\mu = 0$.


\providecommand*\mcitethebibliography{\thebibliography}
\csname @ifundefined\endcsname{endmcitethebibliography}
  {\let\endmcitethebibliography\endthebibliography}{}

\begin{widetext}

\begin{figure}[t]
\centering
\subfigure{\includegraphics[width = 0.36\textwidth]{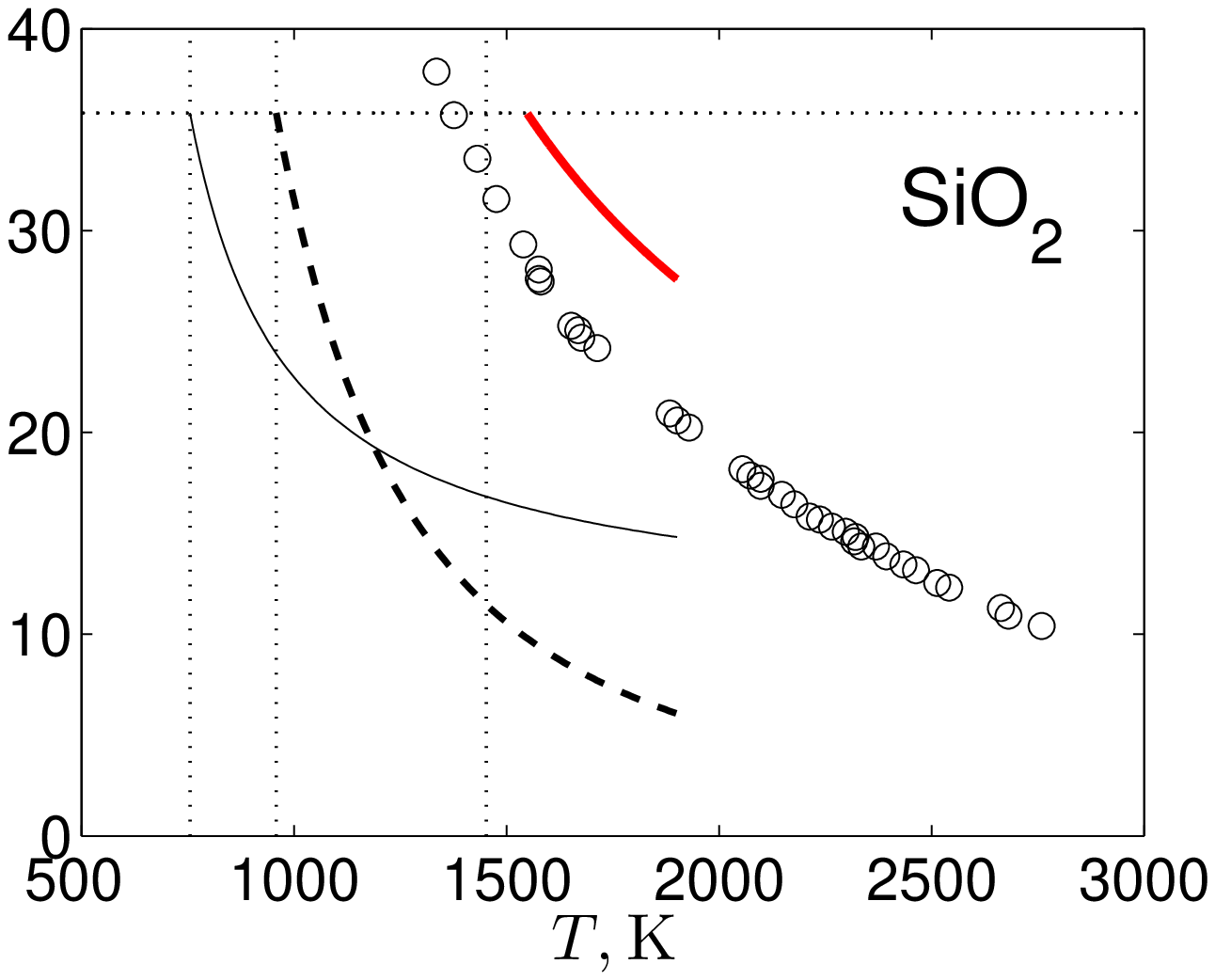}} \hspace{15mm}
\subfigure{\includegraphics[width = 0.36\textwidth]{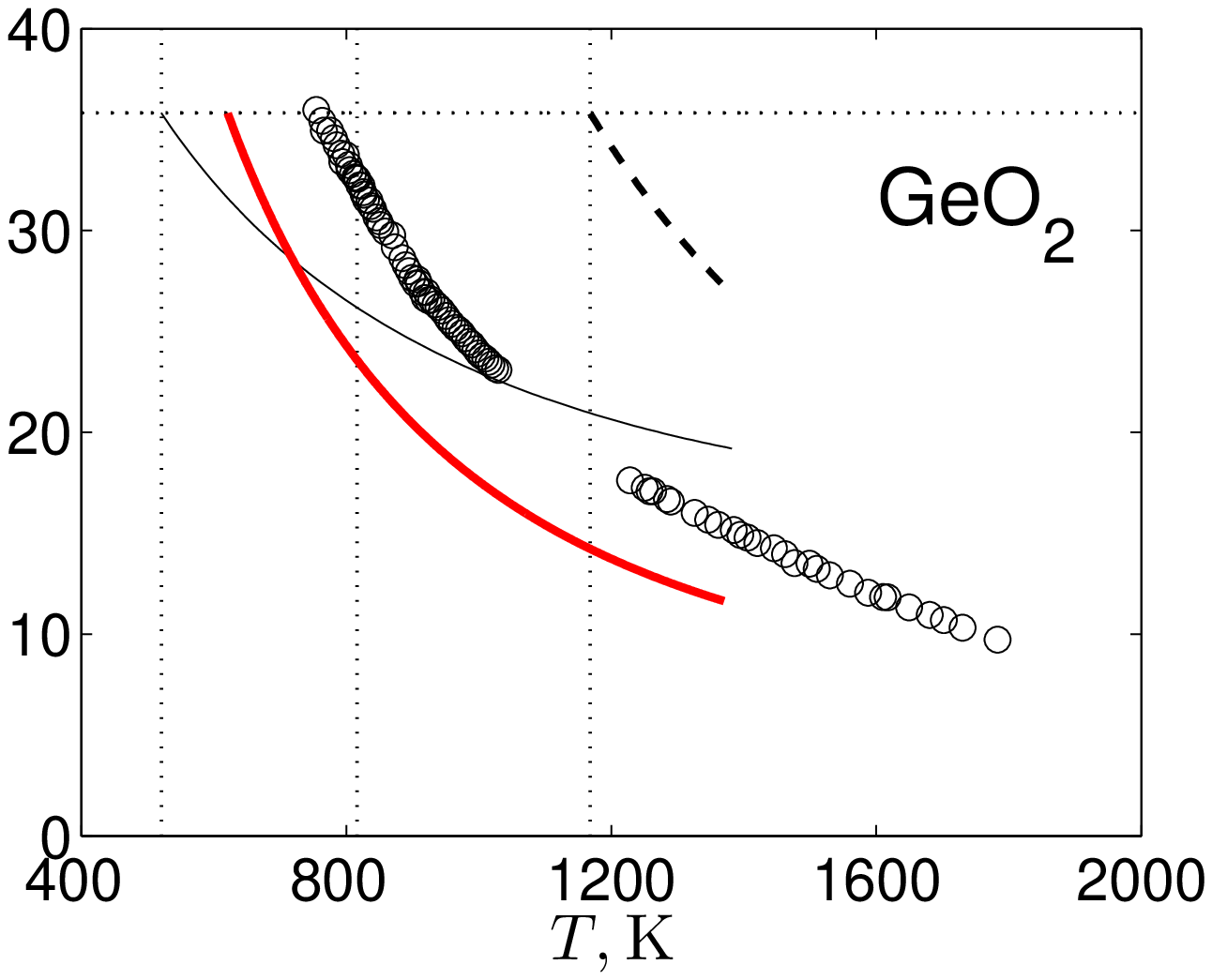}} \\
\subfigure{\includegraphics[width = 0.36\textwidth]{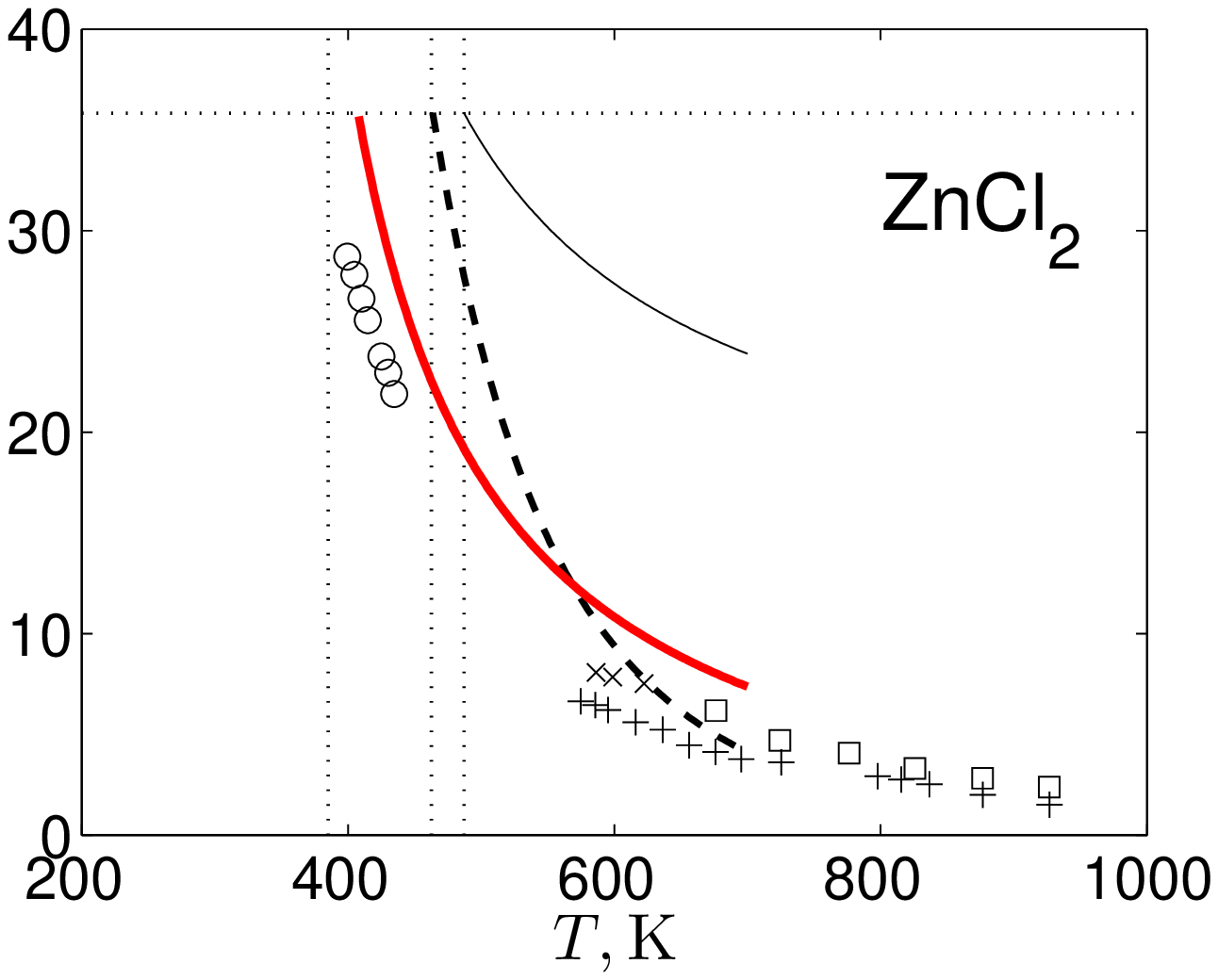}} \hspace{15mm}
\subfigure{\includegraphics[width = 0.36\textwidth]{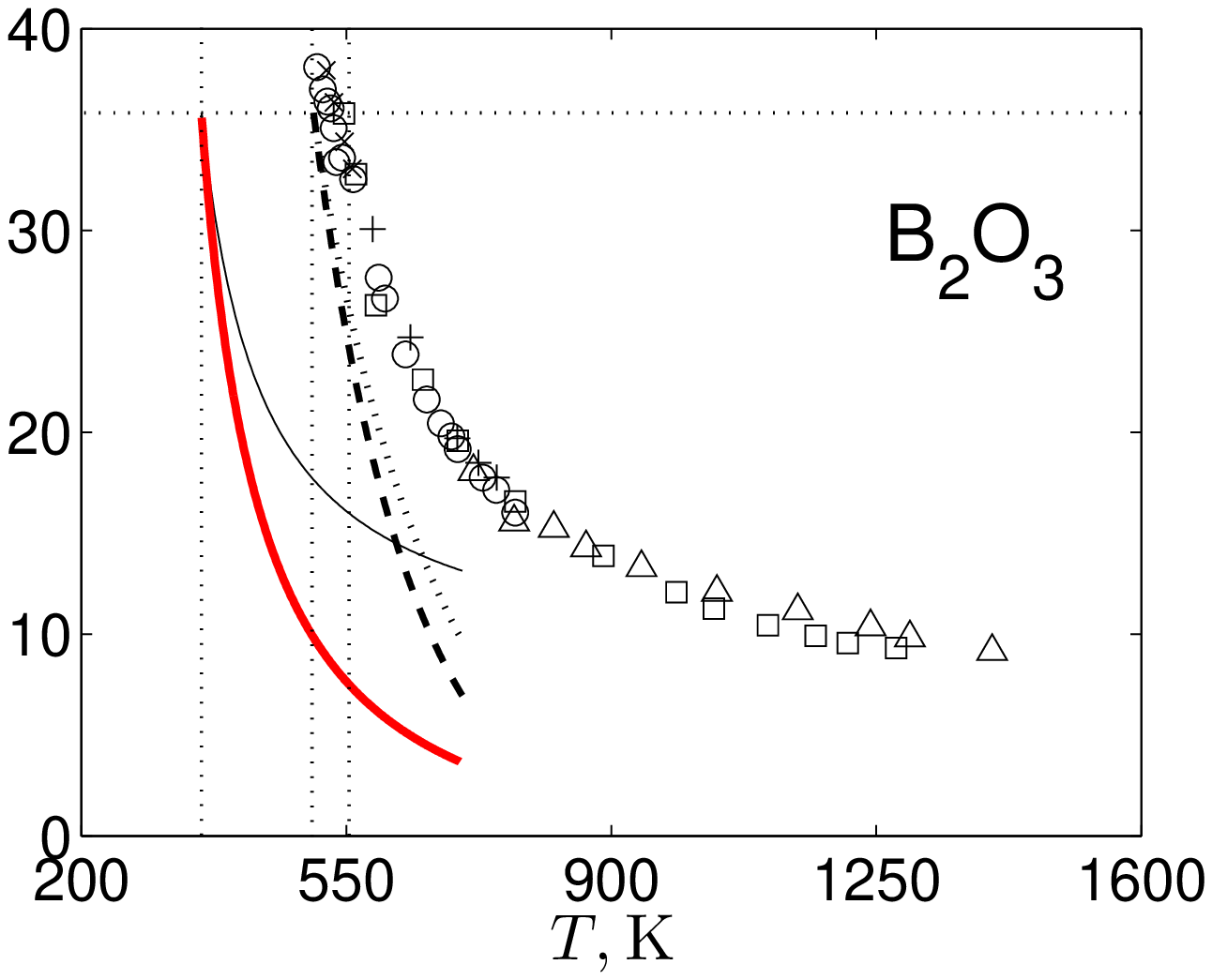}} \\
\subfigure{\includegraphics[width = 0.36\textwidth]{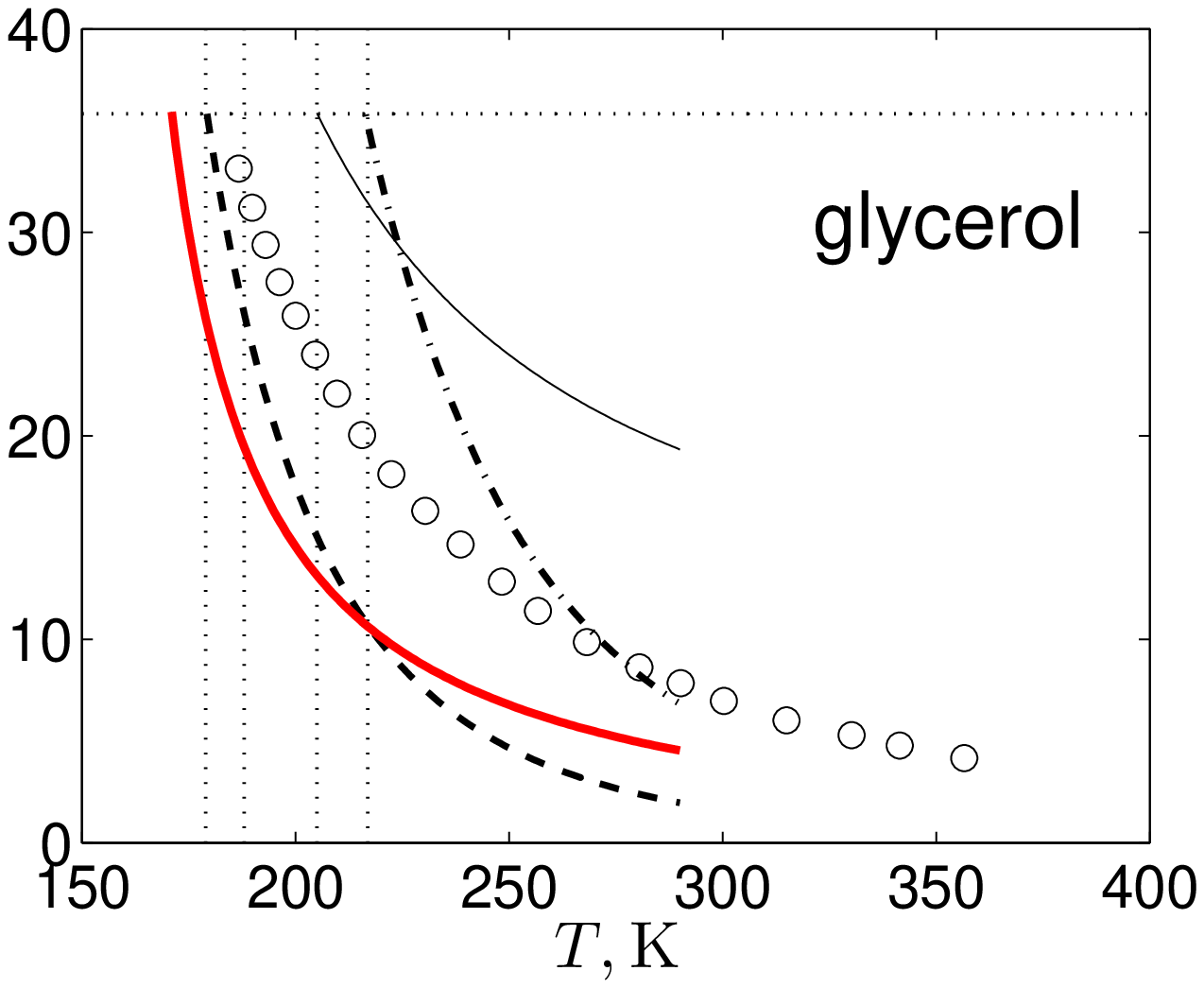}} \hspace{15mm}
\subfigure{\includegraphics[width = 0.36\textwidth]{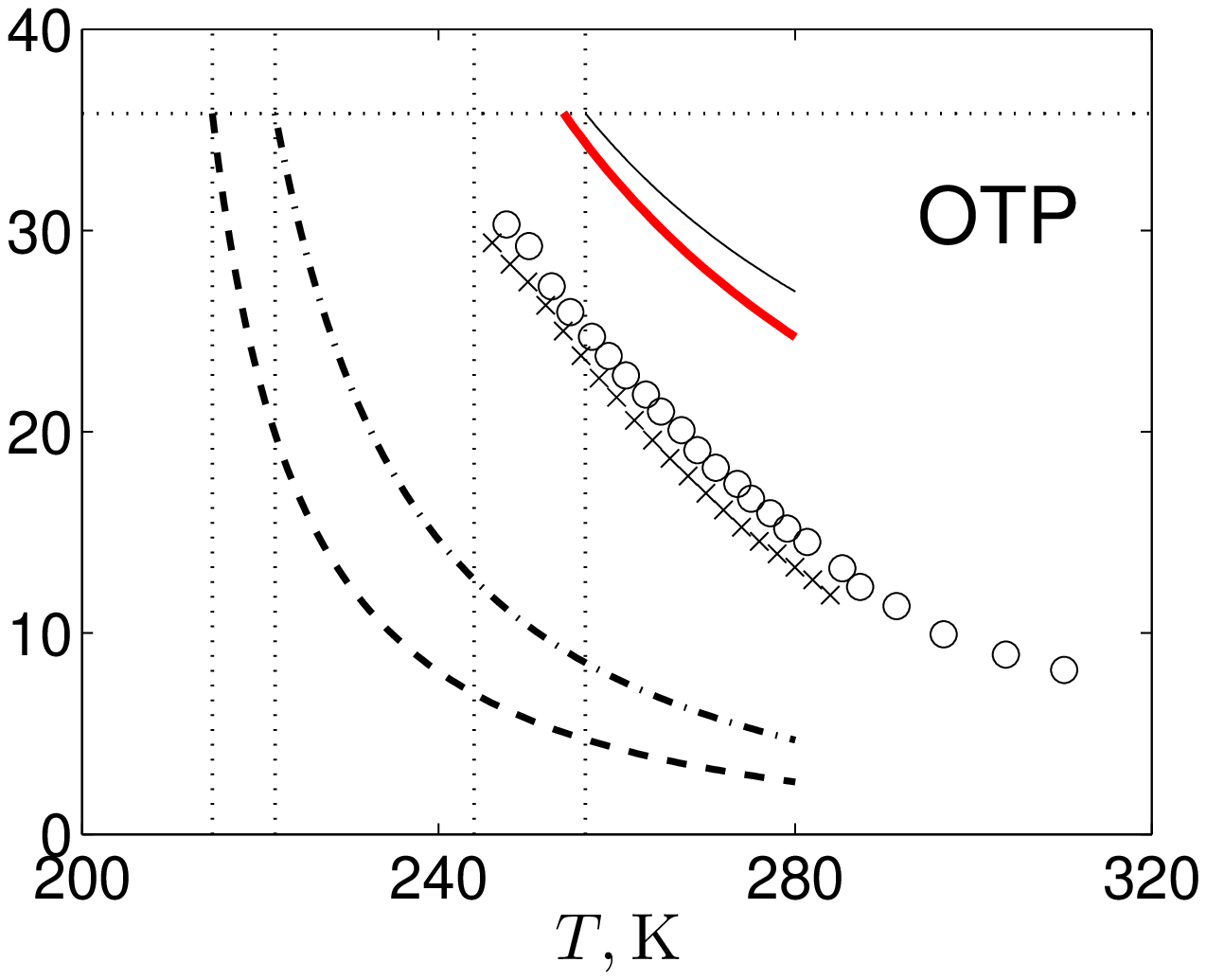}} \\
\subfigure{\includegraphics[width = 0.36\textwidth]{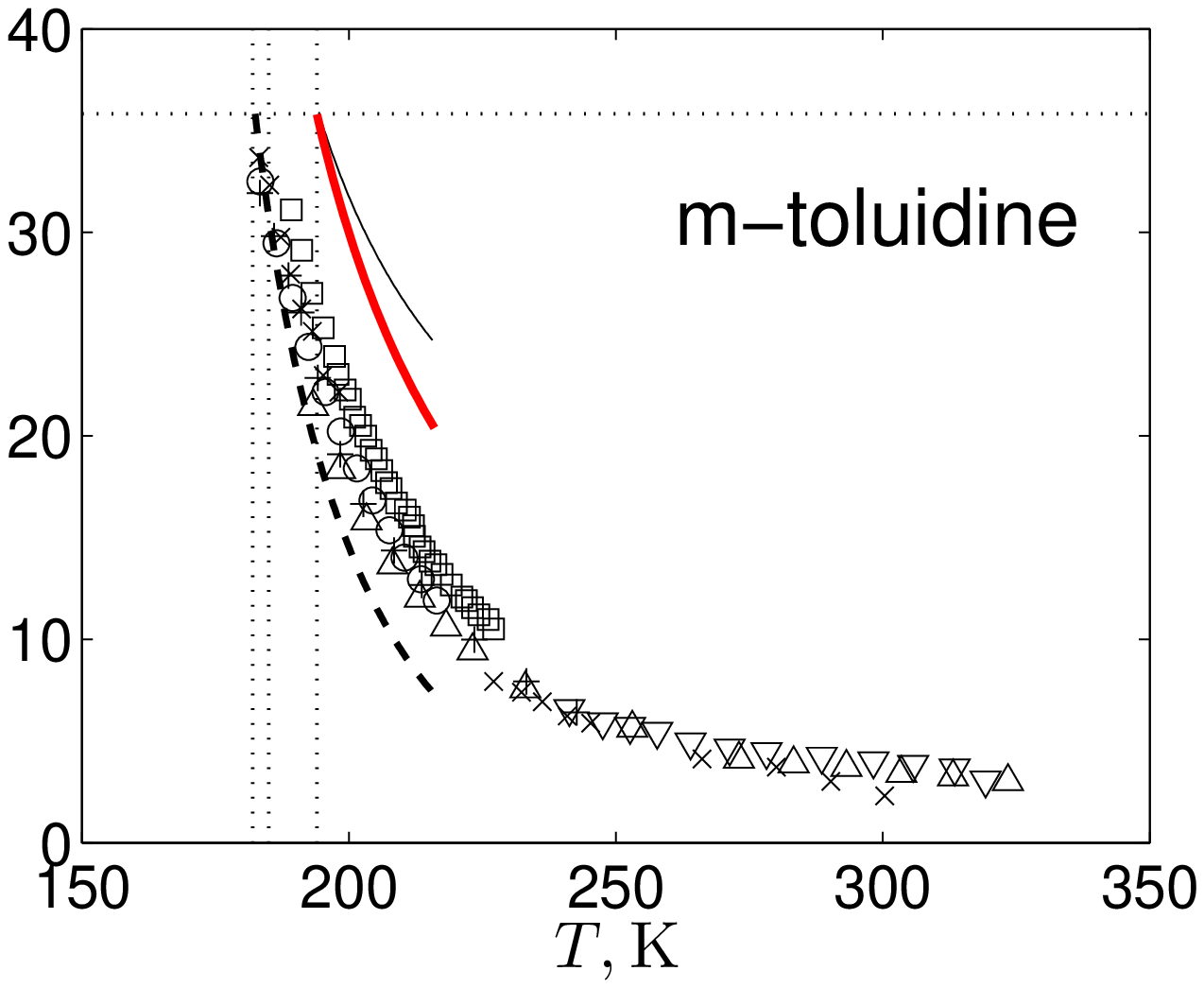}} 
\caption{\label{Cal} The free energy barrier for $\alpha$-relaxation
  (divided by $k_B T$) as a function of temperature calculated using
  three approximations. Two earlier approximations,~\cite{RWLbarrier}
  due to Xia and Wolynes~\cite{XW} (XW) and Rabochiy and
  Lubchenko~\cite{RL_sigma0} (RL) are shown by the thin solid and
  thick dashed line respectively. For these calculations, the
  calorimetric bead count was used.  The present approximation is
  shown with the thick solid red line.  The symbols correspond to
  experimental data; different symbols denote distinct
  experiments. Because temperature dependences of the structure factor
  $S(k)$---which is one of the experimental inputs in the RL
  calculation---are not available, the structure factors measured near
  $T_g$ have been employed. The dash-dotted lines for glycerol and OTP
  correspond to the RL barrier computed using a higher temperature
  $S(k)$.}
\end{figure}

\begin{figure}[t]
\centering
\subfigure{\includegraphics[width = 0.36\textwidth]{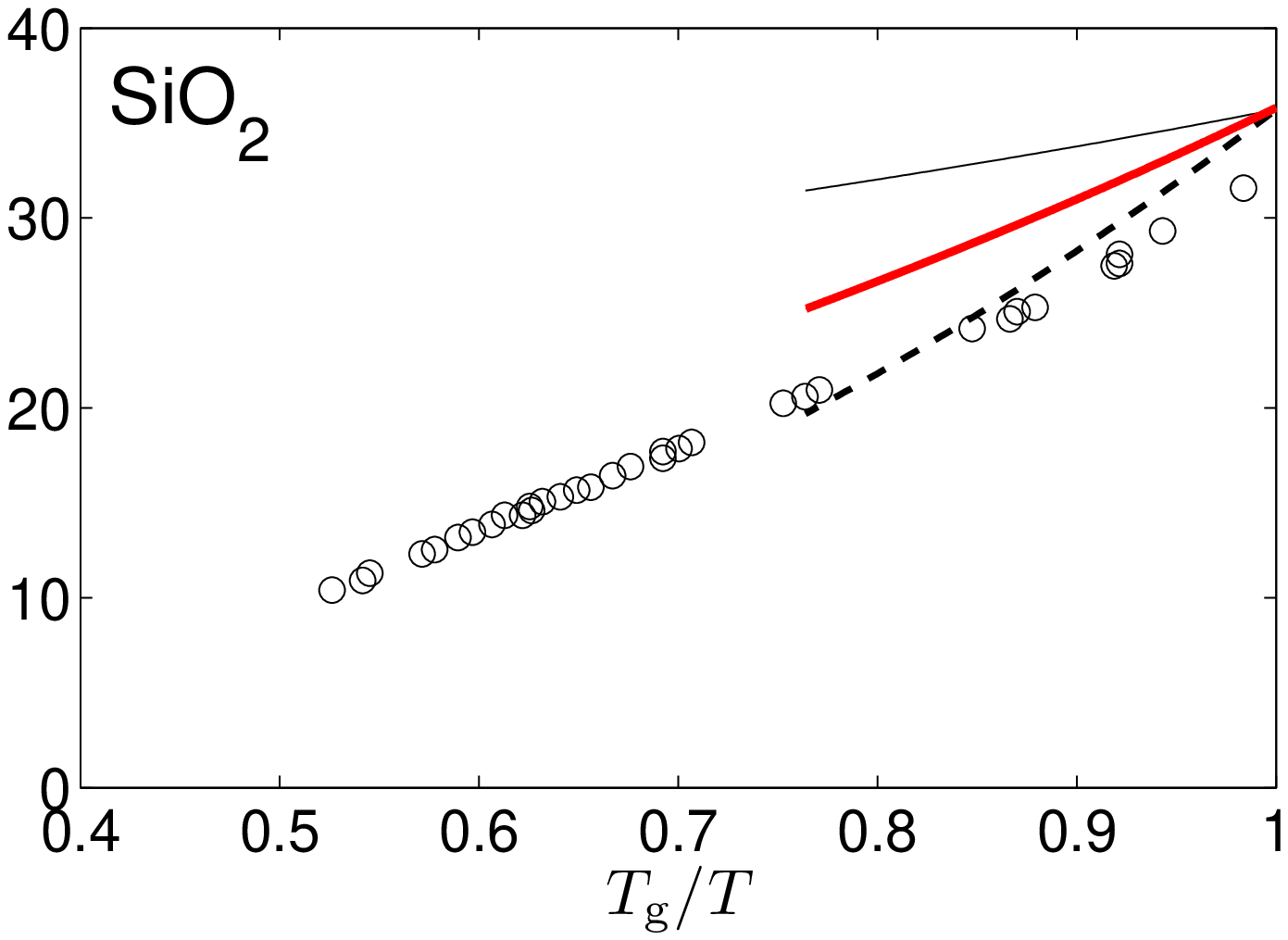}} \hspace{15mm}
\subfigure{\includegraphics[width = 0.36\textwidth]{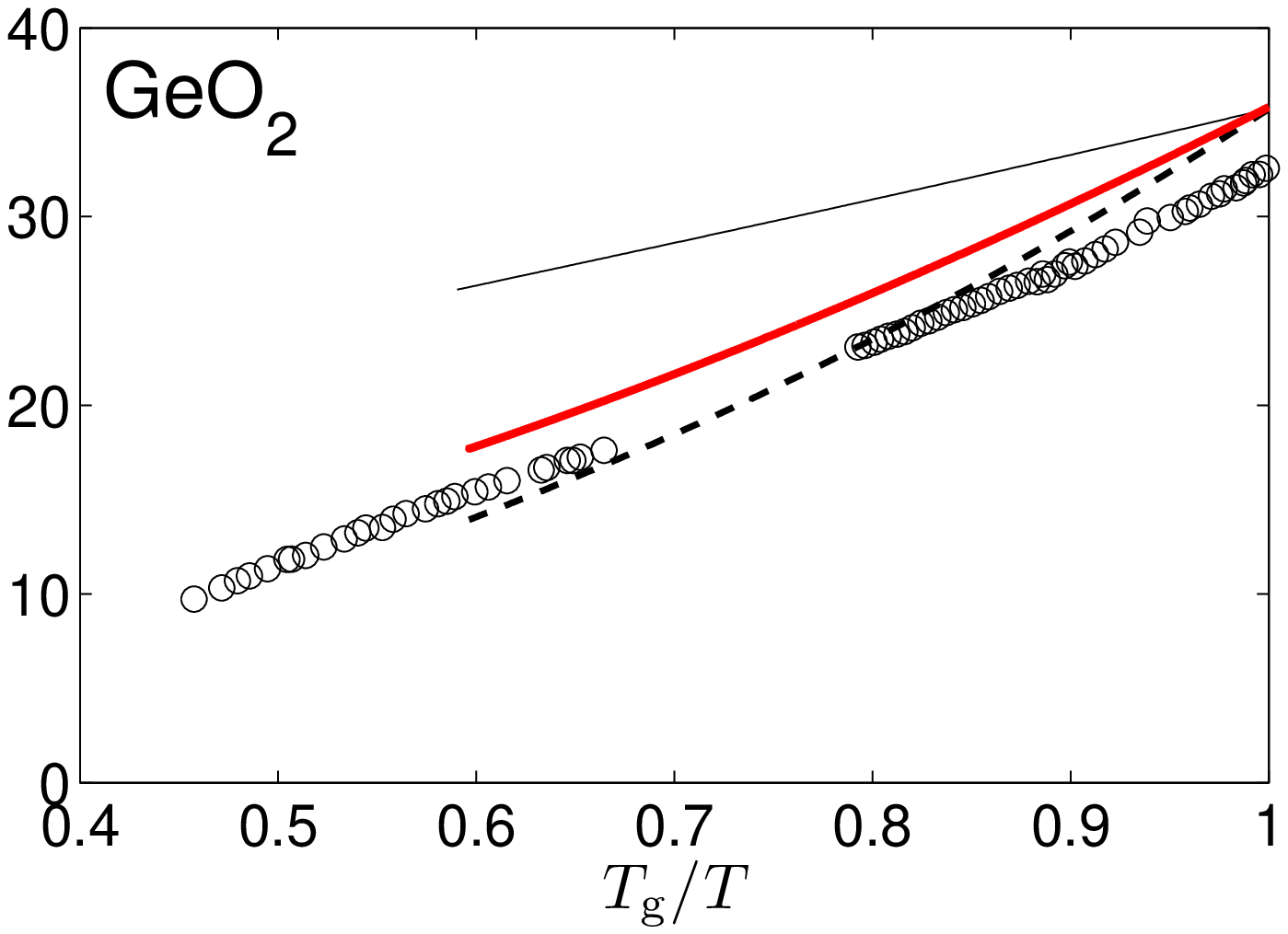}}\\
\subfigure{\includegraphics[width = 0.36\textwidth]{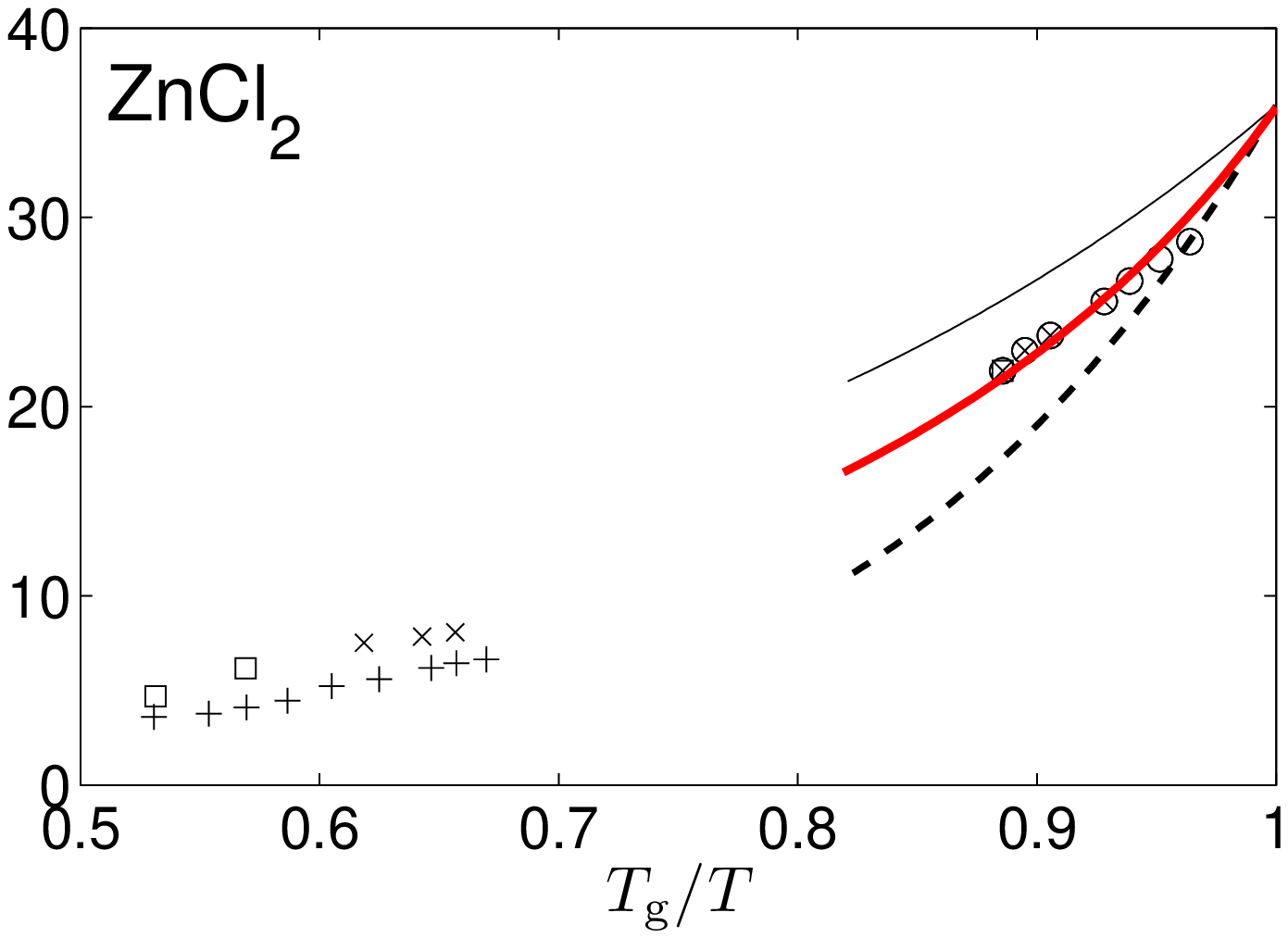}} \hspace{15mm}
\subfigure{\includegraphics[width = 0.36\textwidth]{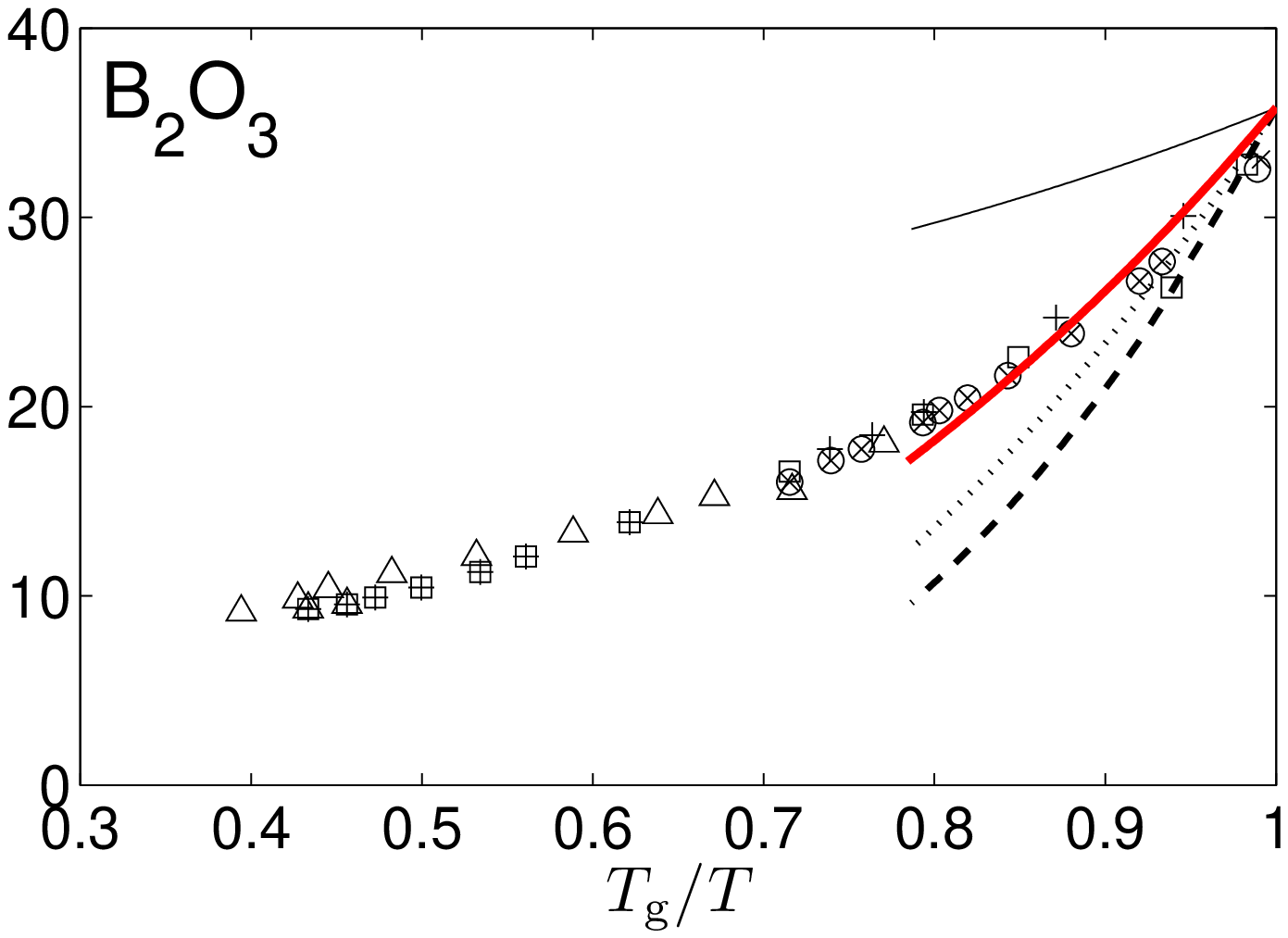}}\\
\subfigure{\includegraphics[width = 0.36\textwidth]{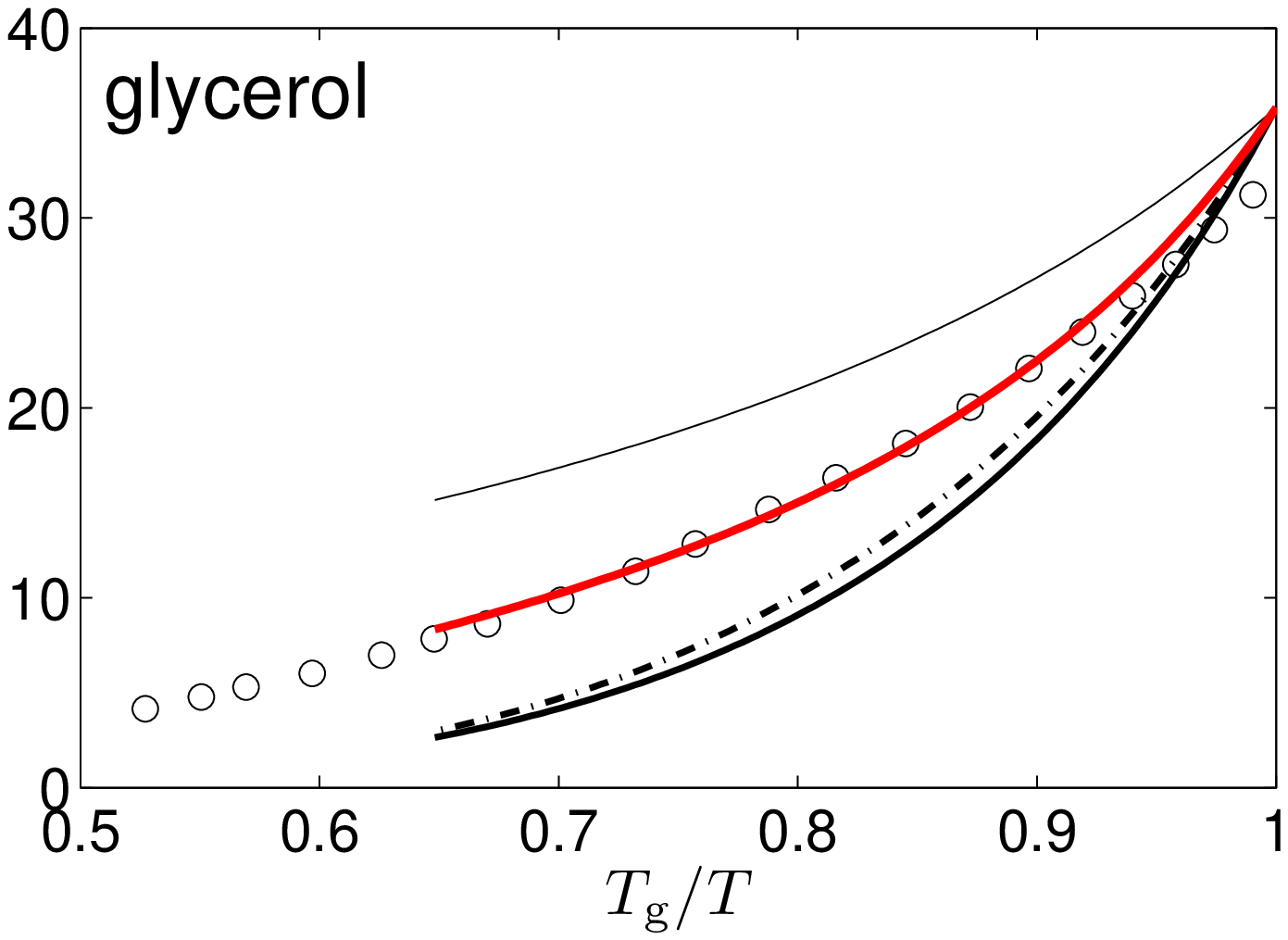}} \hspace{15mm}
\subfigure{\includegraphics[width = 0.36\textwidth]{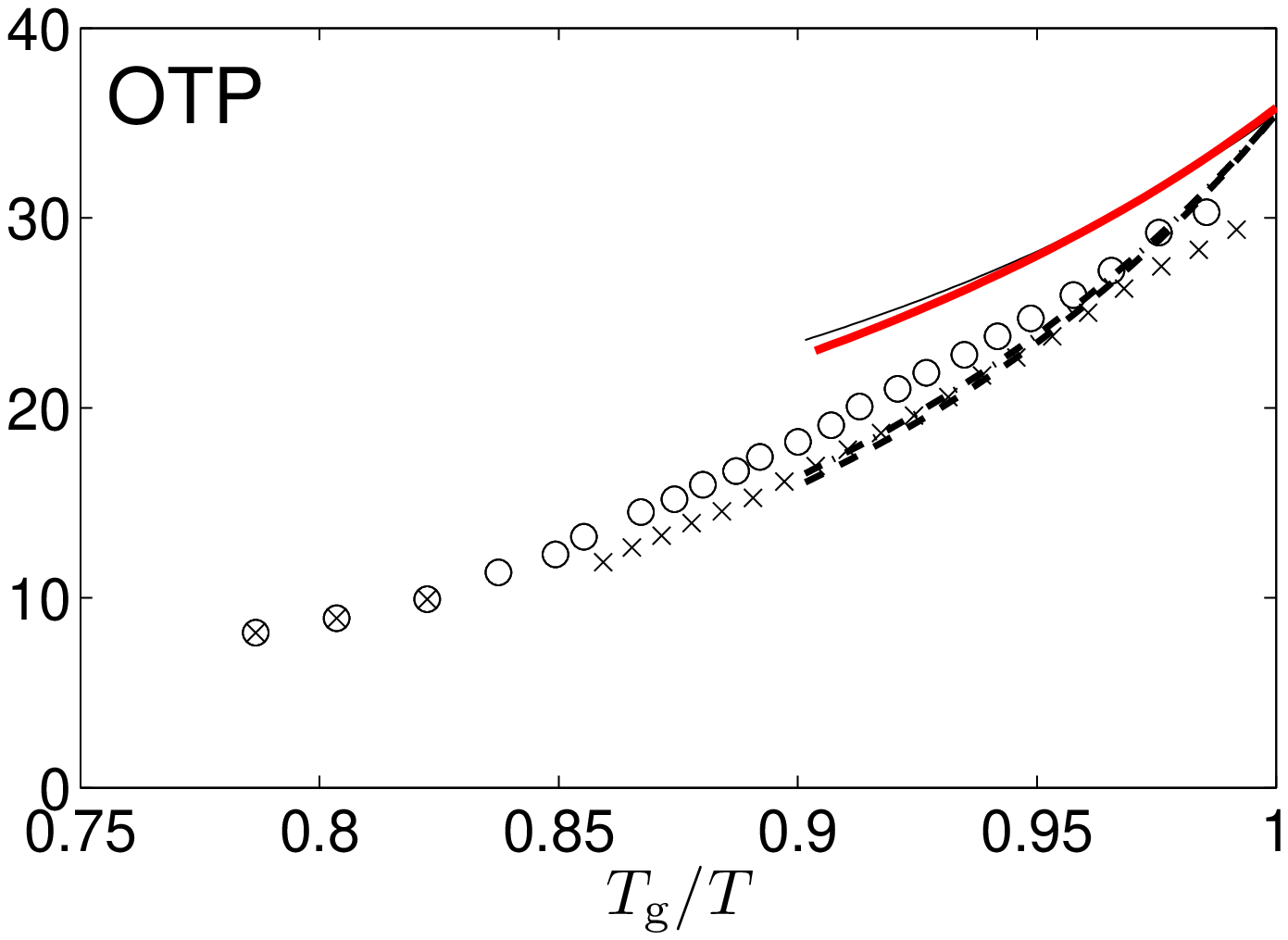}}\\ 
\subfigure{\includegraphics[width = 0.36\textwidth]{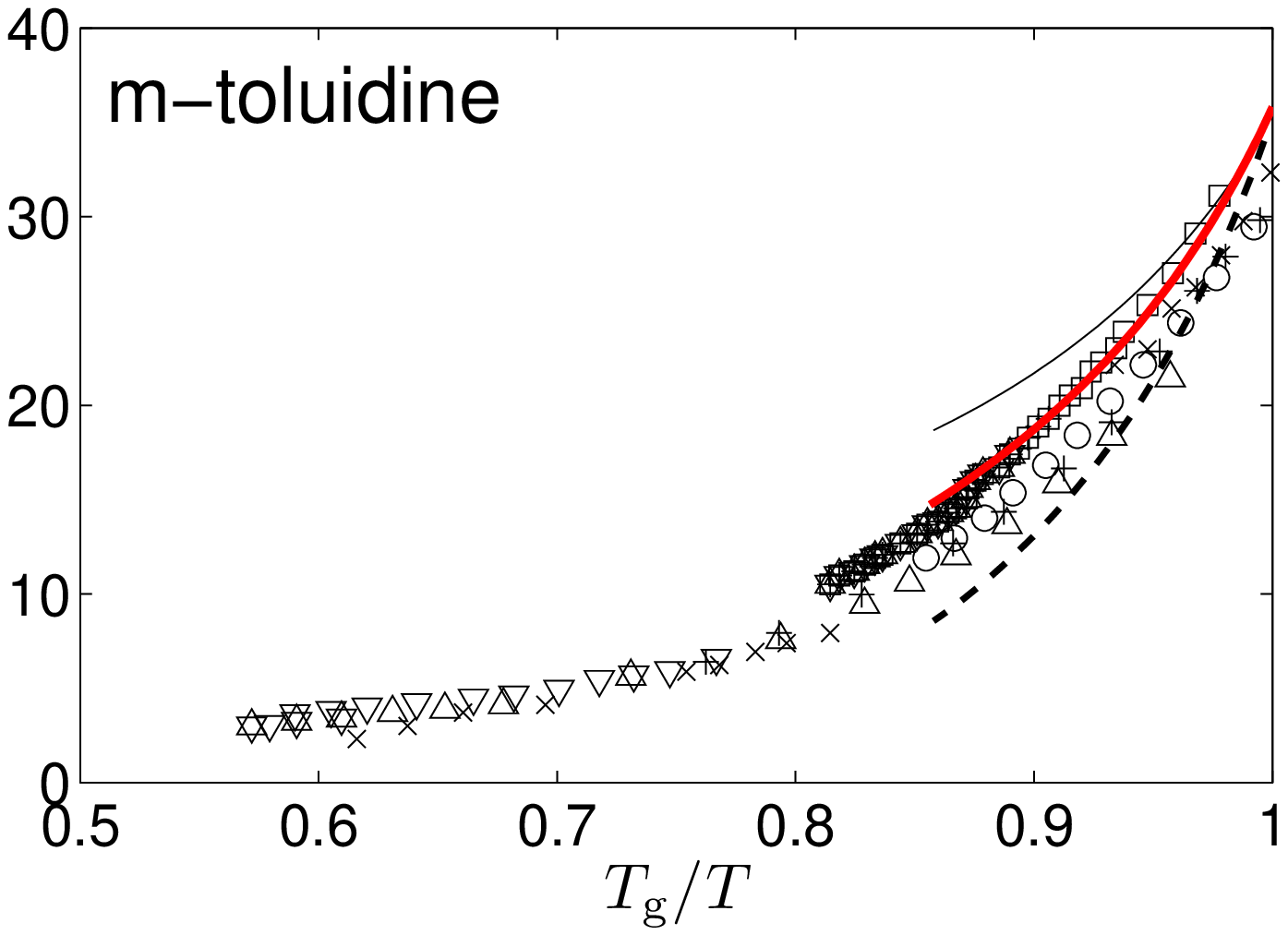}}
\caption{\label{SC} Adjusted barrier for $\alpha$-relaxation (divided
  by $k_B T$) as a function of inverse temperature. The present
  approximation, from Eq.~(\ref{FKsc1}), is multiplied by a constant
  so as to bring the activation exponent at $T_g$,
  $F^\ddagger(T_g)/k_B T_g$, to a fixed value of 35.7 (thick solid red
  line).  In the XW (thin solid line) and RL (thick dashed line)
  approximations, the bead size was determined self-consistently to
  achieve $F^\ddagger(T_g)/k_B T_g = 35.7$.  In all graphs, the
  temperature is scaled by the {\em experimentally} determined
  $T_g$. }
\end{figure}

\end{widetext}

\end{document}